\documentclass[letterpaper]{article}
\usepackage{amsfonts,amssymb,amsmath,bm,bbm,euscript,mathrsfs,calligra}
\usepackage[usenames]{color}
\usepackage{color,soul}
\usepackage[utf8]{inputenc}
\usepackage[T1]{fontenc}
\usepackage{textcomp}
\usepackage{graphicx}
\usepackage[tight,sf]{subfigure}
\usepackage{listings}
\usepackage{hyperref}
\usepackage{bookmark}
\usepackage{authblk}

\DeclareMathAlphabet{\mathcalligra}{T1}{calligra}{m}{n}
\DeclareFontShape{T1}{calligra}{m}{n}{<->s*[2.2]callig15}{}

\newcommand{\bfT}{\mathbf{T}}
\newcommand{\bfL}{\mathbf{L}}

\newcommand{\bft}{\mathbf{t}}
\newcommand{\bfl}{\mathbf{l}}
\newcommand{\bfn}{\mathbf{n}}
\newcommand{\bfu}{\mathbf{u}}
\newcommand{\bfX}{\mathbf{X}}
\newcommand{\bfY}{\mathbf{Y}}
\newcommand{\bfe}{\mathbf{e}}
\newcommand{\bff}{\mathbf{f}}
\newcommand{\rmT}{\mathrm{T}}
\newcommand{\rmL}{\mathrm{L}}
\newcommand{\rmt}{\mathrm{t}}
\newcommand{\rml}{\mathrm{l}}
\newcommand{\calD}{\mathcal{D}}

\newcommand{\calT}{\mathcal{T}}

\begin{document}

\title{\huge {Isometric bending requires local constraints on free edges}}
\author{\Large
Jemal Guven$^1$\footnote{\href{mailto:jemal@nucleares.unam.mx}{jemal@nucleares.unam.mx}}, 
Martin Michael M\"uller$^{2}$, and Pablo V\'azquez-Montejo$^{3}$\footnote{\href{mailto:pablo.vazquez@correo.uady.mx}{pablo.vazquez@correo.uady.mx}}
}
\maketitle

\begin{center}
\textit{$^1$ Instituto de Ciencias Nucleares, Universidad Nacional Autónoma de México\\
 Apdo. Postal 70-543, 04510, Ciudad de México, México}
\\[0.25cm]
\textit{$^{2}$ Laboratoire de Physique et Chimie Th\'eoriques - UMR 7019, Universit\'e de Lorraine\\
1, boulevard Arago, F-57070 Metz, France}
\\[0.25cm]
\textit{$^{3}$ Cátedras Conacyt - Facultad de Matemáticas, Universidad Autónoma de Yucatán\\
Periférico Norte, Tablaje 13615, 97110, M\'erida, Yucat\'an, México}
\end{center}


\begin{abstract}
While the shape equations describing the equilibrium of an unstretchable thin sheet that is free to bend are known, the boundary conditions that supplement these equations on free edges have remained elusive. Intuitively, unstretchability is captured by a constraint on the metric within the bulk. Naively one would then guess that this constraint is enough to ensure that the deformations determining the boundary conditions on these edges respect the isometry constraint. If matters were this simple, unfortunately, it would imply unbalanced torques (as well as forces) along the edge unless manifestly unphysical constraints are met by the boundary geometry. In this paper we identify the source of the problem: not only the local arc-length but also the geodesic curvature need to be constrained explicitly on all free edges. We derive the boundary conditions which follow. Contrary to conventional wisdom, there is no need to introduce boundary layers. This framework is applied to isolated conical defects, both with deficit as well, but more briefly, as surplus angles. Using these boundary conditions, we show that  the lateral tension within a circular cone of fixed radius is equal but opposite to the radial compression,  and independent of the deficit angle itself.  We proceed to examine the effect of an oblique outer edge on this cone perturbatively demonstrating that both the correction to the geometry as well as the stress distribution in the cone kicks in at second order in the eccentricity of the edge. 
 \end{abstract}

\noindent Pacs: 03.50.-z Classical field theories,
46.32.+x Mechanical instability,
87.16.D- Membranes, bilayers, and vesicles
\\\\
Keywords: {Geometric variational principles, isometries}


\section{Introduction}

Consider a thin sheet of paper that is free to deform. If thin enough, it will bend without significant stretching.  What stretching occurs is always localized within a network of point-like defects, connected by ridges, where the stresses within the sheet get focused \cite{Witten,WittenReview}. In this limit, determining the shape of the sheet outside of these defects is a geometrical problem. Inextensibility translates, in geometrical terms, to a constraint on the metric induced on the surface;
the only deformations consistent with the integrity of the sheet must be isometries, fixing this metric (see, for example, \cite{Spivak, Verpoort, Audoly}). Thus, to determine the equilibrium form that the sheet adapts, the bending energy needs to be minimized in a way that accommodates this local constraint. There is a snag: it is not obvious how to implement this constraint in the boundary conditions on any free edges.
\vskip1pc \noindent
A somewhat pessimistic view is that the thin sheet limit is ill-defined at its free edges. Yet, 
hold a sheet of paper in one hand and let it droop gently under gravity. While defects may form within the sheet, it is clear that nothing dramatic is occurring on its free edges.  Conventional wisdom has it that a boundary layer needs to be introduced to treat these edges \cite{WittenReview}.
\vskip1pc \noindent
Something is amiss. Indeed, if one treats the limit cavalierly, imposing isometry only in the bulk, to ones dismay the boundary conditions on the free edges appear to lead to an unphysical conclusion. Specifically torques appear to be unbalanced or, almost as bad, they are balanced only if the 
surface is minimal on this boundary. This would imply that the two principal curvatures vanish there or that the surface must be planar. While a drooping sheet might be coerced to oblige, a circular paper cone certainly does not. In addition, while---as one can check---the integrated force along the full boundary vanishes, force balance fails locally. One might be tempted to agree that the limit is unphysical. However, as we will argue, the problem is not with the limit itself, but how we treat the boundary in the limit: as we will show, satisfying the constraint in the bulk is no guarantee that it will be satisfied on the boundary. We propose a modification of the variational principle, constraining explicitly the boundary variations to be consistent with isometry. This may not be the end of the story, but torques are now balanced and forces vanish pointwise on free boundaries. 
\vskip1pc \noindent
The metric constraint in the bulk is tensorial in nature. This imposes three differential conditions on the deformation vector. If one attempts in an apparently reasonable way to impose all three locally on the boundary curve,  this will
constrain its behavior under deformation in an unphysical way. A signal that this is the case, as we will describe, is the fact that two of the three conditions involve \textit{derivatives} of tangential deformations along the boundary co-normal\footnote{The vector normal to the boundary and tangential to the surface.}, terms that never arise as boundary terms in the variation of any bulk bending energy.
The balance of torques is associated with the freedom to rotate the normal vector freely about the boundary tangent direction, or equivalently, the vanishing of the coefficient of the derivative of the normal deformation along the boundary conormal. No other derivatives of boundary variations should show up.
\vskip1pc \noindent
We argue that the appropriate constraints on the boundary curve are that its arc-length as well as its geodesic curvature---an invariant under isometry---are fixed locally \cite{Okinawa}.  Despite initial appearances, these two constraints are 
\textit{not} equivalent to their three metric counterparts. Technically, the  arc-length constraint is equivalent to the projection of the metric constraint along the boundary tangent; but as we will see, it is not automatically induced on the boundary by the bulk metric constraint, and nor is it redundant.
However the arc-length constraint  fails to produce derivatives along the boundary co-normal of the normal deformation. So it is not enough. The constraint on the boundary geodesic curvature does, under variation, provide the missing term without introducing unphysical rotations;  it is not implied by the boundary constraint on the metric.  These two constraints contribute to all four boundary conditions. In particular, the multiplier constraining the geodesic curvature, as the boundary normal rotates, restores the balance of torques. 
These boundary constraints have a direct impact on the stress distributed within the sheet as well as the forces it transmits across boundaries. We will show that the boundary corrected force vanishes locally on free boundaries, a good sign that we are on the right track.
\vskip1pc \noindent
We will examine a cone (not necessarily circular) to assess how well our proposal fares. This is the simplest defect on a planar sheet.  A cone is characterized intrinsically by its surplus or deficit angle. If this angle vanishes, non-trivial behavior requires external forces. Pomeau and Ben Amar, Cha\"ieb et al, and Cerda and Mahadevan showed that a defect-free planar sheet subjected to a localized force (picture  a circular disc poked at its center into a cup) forms a non-trivial conical defect \cite{PomeauBenAmar97, geminard98, Maha98, Maha05}. If the surplus is negative (a deficit angle), the equilibrium shape in the absence of external forces is a circular paper cone if the radial distance from its apex to its boundary is fixed; however, if set into motion about an axis, such cones will develop non-trivial patterns and non-trivial material motion within the sheet \cite{GHM13}. A few years ago it was also pointed out that the equilibrium shapes of a surplus cone display strikingly interesting behavior compared to their deficit counterparts, even when no external forces are acting.  The equilibria are described by conical monkey saddles with $n$-fold symmetry, $n\ge 2$, their bending energies ordering with $n$ \cite{MBG08}. See also \cite{DBAAlgae}.
The analysis in \cite{MBG08} built on the earlier development of a variational framework to treat bending without stretching \cite{GM08} developed by two of the present authors.
\vskip1pc \noindent
In neither reference \cite{MBG08} nor \cite{GM08} were the boundary conditions on the conical rim consistent with isometry ever fully addressed. This omission was legitimate  due to the coincidence--unfortunate in retrospect--that the conical shape can be determined without any detailed knowledge of the boundary behavior whenever the distance from the apex to the boundary is fixed. In general, however, one is at a complete loss to determine how stress gets distributed within the cone without this information. This is true even in a circular paper cone. Had a more general tangent developable surface been considered, one would have run into the problem straightaway.  Using the modified boundary conditions we will show how to determine the stress  distributed within circular cones. There is no need to invoke boundary layers.  We will show explicitly how the  force transmitted locally across free boundaries vanishes when the new terms are added;
with the naive  boundary conditions this fails, another red flag, if one were needed, that something was amiss.
  \vskip1pc \noindent
The importance of the boundary conditions becomes all the more evident  as soon  as we relax the circular symmetry.  The simplest extension is to replace the circular rim  of the  paper cone by an ellipse.  The cone responds to its altered boundary, which in turn gets deformed. The loss of rotational invariance implies the loss of the conservation law associated with it.   We will show that to determine the shape, never mind the stress, we need to now solve three coupled second-order ordinary differential equations. We will confine ourselves to a perturbative treatment of this problem, not just to keep the focus on the physics, but also to limit the length of the paper. At first order in the boundary ellipticity, there is no change in the bulk circular conical geometry. This accords with observation: make an oblique cut along the rim of a conical paper cup.  There is no perceptible change in the geometry. And nor is there any correction to the distribution of stress within the cone at this order. Significant changes first show up at second order which we will describe in this paper. Curiously, there are symmetry issues that require proceeding to third order to settle.  At this point a numerical treatment is called for.  The essential point having been made, we will leave this for another day.
\\\\
It should be mentioned that there is an enormous and growing literature on the subject of isometric bending.  The twisting of inextensible planar strips was addressed in \cite{Mobius,Starostin}. More recent work has questioned  
the validity of this work (see, for example, \cite{Fried2016,Fried2018})  but the reasons given, certainly in \cite{Fried2016}, are not generally accepted (see, for example, \cite{Healey2019}). A constrained variational problem closely related to the one we consider has been addressed in \cite{Steigmann2018}.
There has also been a growing interest in the  bending of non-Euclidean geometries \cite{Sharon, Efrati11, Gemmer2012, Gemmer2013, Gemmer2016}. It is likely that the boundary problem addressed here, which does not depend explicitly on the flatness of the sheet, is relevant to 
these different directions of research.


\section{Isometry modified shape equations and stresses}

We will suppose that the sheet is thin compared to its radius of curvature. Let it be described as a parametrized surface $(u^1,u^2) \mapsto \mathbf{X}(u^1,u^2)$. The associated bending energy is given by \cite{Helfrich73,Seifert97}
\begin{equation} \label{HB}
H[\mathbf{X}] = \int dA \, \mathcal{H} \,, \quad \mathcal{H} =\frac{1}{2}  K^2  +  \mathrm{k}_G\, K_G  \,,
\end{equation}
where $dA$ is the element of area; ${\cal H}$ is the bending energy density, involving (twice) the mean curvature $K = C_1+C_2$ and the Gaussian curvature $K_G= C_1 C_2$; $C_1$ and $C_2$ are the two principal curvatures. The mean curvature rigidity modulus is set to one; the relative value of the Gaussian rigidity modulus is denoted $\mathrm{k}_G$. We are interested specifically in surfaces that are metrically flat, so that $K_G=0$. Conventionally, however, this term does contribute to boundary conditions so we do not drop it in order to follow its fate. 
\vskip1pc \noindent
The sheet is to a good approximation unstretchable: this translates geometrically into the statement that the induced metric on the surface, given by  $g_{ab}= \mathbf{e}_a\cdot \mathbf{e}_b$ is fixed, where $\mathbf{e}_a=\partial_a\mathbf{X}$, $a=1,2$ are the two tangent vectors adapted to the parametrization by $u^1$ and $u^2$. The unit normal vector will be denoted by ${\bf n}$. 
To determine the equilibrium shapes of the sheet,  the isometry constraint needs to be imposed on deformations of the sheet in the calculus of variations. One way to do this, proposed by  two of the authors, is to impose the constraint using the method of Lagrange multipliers \cite{GM08}. One thus constructs the functional
\begin{equation} \label{VCdef}
H_C[\mathbf{X}] =  H[\mathbf{X}] - \frac{1}{2} \int dA \, T^ {ab} (g_{ab}- g_{ab}^{(0)})\,,
\end{equation}
with the introduction of a symmetric tensor-valued Lagrange multiplier $T^{ab}$. $T^{ab}$ constrains the metric to coincide with some given metric $g_{ab}^{(0)}$. 
\vskip1pc \noindent
Consider a deformation of the surface, $\mathbf{X}\to \mathbf{X}+\delta \mathbf{X}$, and let us decompose 
$\delta \mathbf{X}$ into its tangential and normal parts,
\begin{equation} \label{delXtn}
\delta\mathbf{X}= \Psi^a \mathbf{e}_a + \Phi\mathbf{n}\,.
\end{equation} 
It is straightforward to show--using methods developed elsewhere (in \cite{GM08}, and, for example, \cite{CGS03})--that the first variation of the bulk constrained Hamiltonian  under the surface deformation described by the vector $\delta\mathbf{X}$, can be cast as
\begin{equation} \label{del HC}
\delta H_C[\mathbf{X}] = \int dA \, 
\left[({\cal E}_0 - T^{ab} K_{ab}) \Phi + \nabla_a T^{ab} \Psi_b \right] + {\sf BT}\,,
\end{equation}
where ${\sf BT}$ denotes the boundary integral, $\nabla_a$ is the covariant derivative compatible with $g_{ab}$ and $K_{ab}= \mathbf{e}_a \cdot \partial_b \mathbf {n}$ is the extrinsic curvature tensor on the surface, with trace $K$  ($K= g^{ab} K_{ab}$). Here ${\cal E}_0$ is given (for a metrically flat surface, $K_G=0$) by
\begin{equation}
{\cal E}_0 = - \nabla^2 K - K \left(\frac{K^2}{2} - 2 K_G\right)\,.
\end{equation}
The crucial boundary integral BT, linear in $\delta\mathbf{X}$, will be discussed  in detail in Sec. \ref{Sec:FBC}. The bulk shape equations, following from Eq.(\ref{del HC}),  are given by
\begin{subequations} \label{EL}
\begin{eqnarray}
{\cal E}_0 - K_{ab} T^{ab} &=& 0 \,;  \label{EL:normal}\\
\nabla_a T^{ab} &=& 0 \,. \label{EL:tangential}
\end{eqnarray}
\end{subequations}
Unstretchability introduces an inhomogeneous and anisotropic additional tangential stress within the sheet identified with the tensor-valued multiplier imposing this constraint.
It is covariantly conserved when the forces are in equilibrium. This stress in turn provides a source term in the shape equation. 
\vskip1pc \noindent
In the presence of free-boundaries this is not the complete story, a fact  the authors of Ref. \cite{GM08} were a little tight-lipped about in their initial proposal. Forward eleven years: we will be somewhat more forthcoming.
\vskip1pc
To better understand the structure of the boundary terms appearing in the variational principle, 
it is useful to know that the first variation (\ref{del HC}) can also be cast in the form 
\begin{eqnarray} 
\delta H_C[\mathbf{X}] = \int dA\, 
(\nabla_a \, \mathbf{f}^a ) \cdot\delta \mathbf{X} + {\sf BT} \,.\nonumber
\end{eqnarray}
The EL derivative of $H$ with respect to ${\bf X}$ can then be cast as a divergence \cite{CG02,JG04}(reviewed in \cite{Guven2018})
\begin{equation} \label{eq:nabf}
\frac{\delta H_C}{\delta {\bf X}} =\nabla_a {\bf f}^a\,, 
\end{equation}
where ${\bf f}^a = {\bf f}_B^a + T^{ab} \mathbf{e}_b $, with a contribution associated with bending given by  (see also \cite{Jenkins1977,Steigmann1999})
\begin{equation} \label{fB}
{\bf f}_B^a =  K( K^{ab}  - \frac{1}{2} K g^{ab} ) \, {\bf e}_b - \nabla^a K \, {\bf n}\,.
\end{equation}
Notice that $\mathbf{f}_B^a$ is  independent of $\mathrm{k}_G$.
\vskip1pc \noindent
In equilibrium, ${\bf f}^a$ is covariantly conserved on the surface or
\begin{equation} \label{nabfr}
\nabla_a \, \mathbf{f}^a = 0 \,.
\end{equation}
The normal and tangential projections of Eq. (\ref{nabfr}) reproduce Eqs. (\ref{EL}a) and (\ref{EL}b). Equation (\ref{nabfr}) is the conservation law associated with translational invariance; physically ${\bf f}^a = f^{ab} \mathbf{e}_b + f^a \mathbf{n}$ is interpreted as the stress tensor, a direct consequence of Noether's theorem. Note that the bending contribution to $f^{ab}$ is quadratic in $K^{ab}$; as such the principal tangential bending stresses will occur along the principal directions of the surface geometry; however, to the bending stress is added the tangential stress associated with the isometric constraint $T^{ab}$ which does not necessarily commute with $K^{ab}$ (unless trivially when the symmetry commands it).
A consequence is that shearing stresses are possible with respect to these directions as we will see in section \ref{section:perp}.


\section{Free boundary conditions} \label{Sec:FBC}

We now look at the boundary conditions informed by this re-expression of the variation in terms of stresses. The arc-length parametrized boundary curve $\ell \mapsto (U^1(\ell),U^2(\ell))$ is identified as a space curve under the composition of maps, $\ell \to \bfY(\ell) = \bfX(U^1(\ell),U^2(\ell))$, reporting surface isometries through $\bfX$. It need not be connected. Let $\bfT =  \dot{\bfY}$ be the unit tangent vector to the boundary curve, where the dot denotes a derivative with respect to the boundary arc length: $\dot{}=\partial_\ell$. We construct a Darboux frame adapted to the boundary, with the tangent $\mathbf{T}$, the \textit{conormal} $\mathbf{L}=\bfn \times \bfT$, and the normal unit vector $\bfn$. Both $\bfT$ and the conormal $\bfL$ can be expanded with respect to the surface adapted tangent vectors: $\bfT = \rmT^a \bfe_a$, $\bfL = \rmL^a \bfe_a$. The surface covector $\rmL_a$ appearing in Eq.(\ref{delHCB}) is given by $g_{ab} \rmL^b$.\footnote{The normalization of $\bfT$ and $\bfL$ translate as $g_{ab} \rmT^a \rmT^b = 1 = g_{ab} \rmL^a \rmL^b$ and $g_{ab} \rmL^a \rmT^b=0$; the latter can be rewritten as $\rmL_a \rmT^a=0$. Notice that the condition $\rmL_a \rmT^a$ is invariant under isometry ($\rmT^a = \dot{U}^a$).} The contribution to the variation of the energy due to surface boundary changes is given by (see, for example, \cite{Guven2018})
\begin{equation} \label{delHCB} 
\delta H_{C\,{\sf boundary}} = - \int d\ell \, \rmL_a \left(\bff^a \cdot \delta \bfX - H^{ab} \bfe_b \cdot \delta \bfn \right)\,.
\end{equation}
In this expression
\begin{equation}
\label{Habdef}
H^{ab} = K g^{ab} + \mathrm{k}_G (K g^{ab} - K^{ab})\,.
\end{equation}
The \textit{force} per unit length transmitted across this line element is given by the projection of the stress tensor \cite{Guven2018}\footnote{Whereas this is the force across interior curves, it is not the force transmitted across the free boundary, which should vanish. As we will see,  it will vanish when boundary constraints accommodating isometry on the boundary are  added.}
\begin{equation} \label{def:fperp}
{\bf f}_\perp := \mathrm{L}_a \mathbf{f}^a = f_{\perp \parallel} \mathbf{T} + f_{\perp \perp} \mathbf{L} + f_\perp {\bf n} \,,
\end{equation}
where the projected quantities onto the Darboux frame are defined by $f_{\perp\parallel} = \mathrm{T}_a \mathrm{L}_b f^{ab}$, $f_{\perp\perp} = \mathrm{L}_a \mathrm{L}_b f^{ab}$ and $f_\perp = \mathrm{L}_a f^a$.
\vskip1pc \noindent
We are interested in casting the boundary variation in terms of geometrical quantities associated with this frame. The equations describing the rotation of this frame as the boundary curve is followed are given by \cite{doCarmo}
\begin{equation} \label{BndyDbxeqs}
\dot{\bfT} = \kappa_g \bfL - \kappa_n \bfn \,, \quad
\dot{\bfL} = -\kappa_g \bfT + \tau_g \bfn \,,\quad
\dot{\bfn} = \kappa_n \bfT - \tau_g \bfL \,.
\end{equation}
The geodesic curvature $\kappa_g$ involves acceleration and thus two derivatives along the curve, whereas the normal curvature $\kappa_n$, and the geodesic torsion $\tau_g$, depend only on the tangent vector, and thus involve a single derivative. While $\kappa_g$  is a measure of curvature tangential---and thus intrinsic---to the surface, $\kappa_n$ is the surface curvature along the direction, $\mathbf{t}$. It is straightforward to show that $\kappa_g= \rmL_a \rmT^b \nabla_b \rmT^a$, whereas $\kappa_n := K_{\parallel \parallel} = K_{ab} \rmT^a \rmT^b$ and $\tau_g : = -K_{\parallel \perp} = - K_{ab} \rmL^a \rmT^b$.
\vskip1pc \noindent
We now decompose the boundary deformation $\delta \mathbf{X}$ with respect to the Darboux frame $\{\bfT,\bfL,\bfn \}$,
\begin{equation} \label{delalphaeulelonsurf}
\delta \bfX = \Psi_\parallel \bfT + \Psi_\perp \bfL + \Phi\, \bfn \,.
\end{equation}
In terms of the components of $\delta \mathbf{X}$, Eq.(\ref{delHCB}) can now be recast as (details are presented in \ref{App:dHbdry})
\begin{equation} \label{boundaryvar}
\delta H_{C\,{\sf boundary}} = -\int d\ell \, \left[  T_{\perp \|}^{\calD}\, \Psi_\parallel + (T_{\perp\perp}^{\calD}- {\cal H} )\, \Psi_\perp- (\nabla_\perp K +\mathrm{k}_G  \dot{\tau}_g)\, \Phi + (K + \mathrm{k}_G \kappa_n)\, \nabla_\perp \Phi \right] \,,
\end{equation}
where $\mathcal{H}$ is the bending energy density defined in Eq.(\ref{HB}); $T_{\perp \parallel}^{\calD} = \rmL_a \rmT_b \, T^{ab}$ and $T_{\perp\perp}^{\calD} = \rmL_a \rmL_b \, T^{ab}$; $\nabla_\perp$ is the derivative along the direction $\bfL$ ($\nabla_\perp = \rmL^a \nabla_a$).\footnote{One integration by parts on $\ell$ was performed to identify the coefficient of $\Phi$.} We introduce the superscript $\calD$ on the components $T_{\perp \perp}^{\calD}$ and $T_{\parallel \perp}^{\calD}$ to indicate that the projections refer to the Darboux frame ($\calD$) adapted to the boundary curve. These generally do not coincide with the projections onto the principal directions of $K^a{}_b$.
\vskip1pc \noindent
Notice that the bending contribution to the term proportional to $\Psi_\perp$ is simply proportional to the unconstrained bending energy density, ${\cal H}$, a feature that is a direct consequence of the reparametrization invariance of $H$ (or any other physically meaningful energy) and the identification of tangential deformations with reparametrizations for any functional that depends only on $\mathbf{X}$. Specifically, if $\delta_{\bm{\Psi}} \mathbf{X}=\Psi^a\mathbf{e}_a= \Psi^a \partial_a \mathbf{X}$, then $\delta_{\bm{\Psi}} H= \int dA \,\nabla_a (\mathcal{H} \Psi^a) = \int d\ell\,\mathcal{H}  \Psi_\perp$.
If it were not for the isometry constraint, which spoils the identification of tangential deformations with reparametrizations, the boundary integrand proportional to $\Psi_\parallel$ would vanish. 
\vskip1pc \noindent
Naively, $\Psi_\parallel$,  $\Psi_\perp$, $\Phi$ as well as $\nabla_\perp\Phi$ can be varied independently on free boundaries. The balance of torques, captured by the freedom to rotate the normal $\mathbf{n}$ on the boundary,  is captured by the vanishing of the coefficient of $\nabla_\perp \Phi$ at each point along the free boundary. If everything were correct this would imply that 
\begin{equation} \label{BCnabPhi}
 K + \mathrm{k}_G \kappa_n = 0 \,.
\end{equation}
One does not have to think very hard, however, to notice that there is something untoward in these boundary conditions: the Gaussian curvature vanishes on a flat sheet so why would it wind up in the boundary conditions; and, if it is ignored, we are led ineluctably to the patently false conclusion that $K=0$ on the boundary.  Together with the vanishing of $K_G$, this would imply that the sheet must be planar there. Something is amiss, the challenge is to discover what is wrong and to fix it.


\subsection{Isometry and surface curves}

The three components of $\delta \bfX$ on free boundaries are not independent. We have just shown that this interdependence is not captured by the multipliers imposing the bulk constraint on the metric appearing in the boundary conditions for this would imply unbalanced torques on free boundaries.  They are not sufficient.  
\vskip1pc \noindent
An apparently reasonable solution would be to impose the linearized bulk isometry constraint, $\delta_\bfX g_{ab}=0$, explicitly on the boundary variations. Using the definition of the induced metric, these three equations can be cast as
\begin{equation}
\bfe_a \cdot \partial_b \delta \bfX + \bfe_b \cdot \partial_a \delta \bfX = 0 \,.
\end{equation}
These equations, in turn, can be decomposed with respect to the projections of $\delta \bfX$ defined by Eq. (\ref{delXtn}):
\begin{equation} \label{delg}
\nabla_a \Psi_b + \nabla_b \Psi_a + 2 K_{ab} \Phi =0\,.
\end{equation}
Typically, we cannot solve these equations analytically. But even if we could, reasonable as they may appear as constraints on the boundary variations, this cannot be right as we will now argue.
\vskip1pc \noindent
The projections along  $\bfT$ and $\bfL$ of Eqs. (\ref{delg}) along the boundary can be cast in the form
\begin{subequations} \label{delgproj}
\begin{eqnarray}
\dot{\Psi}_\parallel - \kappa_g \Psi_\perp  + \kappa_n \Phi &=&0 \,; \label{delgproj1} \\
\nabla_\perp \Psi_\parallel + \dot{\Psi}_\perp - \kappa_g \Psi_\parallel - \kappa_{g \perp} \Psi_\perp - 2 \tau_g\, \Phi &=& 0 \,; \label{delgproj2} \\
\nabla_\perp \Psi_\perp - \kappa_{g \perp} \Psi_\parallel + \kappa_{n \perp} \Phi &=& 0\,, \label{delgproj3}
\end{eqnarray} 
\end{subequations}
where we use the definitions introduced in Eq.(\ref{BndyDbxeqs}) and introduce a few more. Here
$ \kappa_{g \perp}$ is the geodesic curvature along the direction $\bfL$: $ \kappa_{g \perp} =
\rmT_a \rmL^b\nabla_b \rmL^a$, involving a simple interchange of $\bfT$ and $\bfL$ with respect to the definition of $\kappa_g$. We have also used the completeness relationship $\rmT^a \rmT^b + \rmL^a \rmL^b = g^{ab}$  to express $K = \kappa_n + \kappa_{n \perp}$, where $\kappa_{n \perp}$ is the normal curvature along $\bfL$: $\kappa_{n \perp} = K_{ab} \rmL^a \rmL^b$.  Finally, we have used the identity of geodesic torsions along orthogonal curves, $\tau_{g\,\perp} = -K_{ab} \rmT^a \rmL^b = \tau_g$.
\vskip1pc \noindent
The important point to note here is that both $\nabla_\perp \Psi_\parallel$ and $\nabla_\perp \Psi_\perp$ appear in these equations whereas $\nabla_\perp\Phi$ does not. But whereas the latter appears in the  boundary term (\ref{boundaryvar}) arising from the variation of the bulk energy, the first two do not. In general, this suggests that this is not the correct approach to constraining the boundary variations. If we were stubborn and insisted, we would need to include $\nabla_\perp \Psi_\parallel$ and $\nabla_\perp \Psi_\perp$ as independent variations. But because these variations appear nowhere else, the multipliers enforcing Eqs. (\ref{delgproj2}) and (\ref{delgproj3}) would vanish. Only one of the three constraints can be relevant. Not surprisingly,  Eq. (\ref {delgproj}a) is the statement that arc-length is preserved locally along the boundary. The non-appearance of the term proportional to $\nabla_\perp\Phi$ implies that this is  not a sufficient constraint on the boundary variations. 
\vskip1pc \noindent
We propose instead that the relevant constraints on isometric boundary variations involve only the intrinsic geometry of the boundary curve itself as it is embedded in the surface. The obvious 
measure is the local arc-length. Its invariance under isometry is imposed by introducing a local Lagrange multiplier ${\cal T}$, adding to $H_C$ a term  
\begin{equation}
H_1 = \int dv \, {\cal T} (v)  \left( \sqrt{G} - \sqrt{G^{(0)}}\,\right) \,.
\end{equation}
For technical purposes we introduced the one-dimensional induced metric along the boundary when it is parametrized by some external fixed parameter $v$, rather than arc-length $\ell$, so that $d\ell = dv\, \sqrt{G}$, where $G= g_{ab} \tau^a \tau^ b$, with $\tau^a= dU^a/dv$ (not to confused with the geodesic torsion), so that the unit tangent vector is given by $\rmT^a =\tau^a/\sqrt{G}$. ${G}^{(0)}$ is the fixed  metric. 
\vskip1pc \noindent
Now, under a surface deformation, $\delta_\bfX \sqrt{G} = \frac{1}{2}\, \sqrt{G} \, \rmT^a \rmT^b \, \delta_\bfX g_{ab}$, so that  
\begin{equation} \label{delS}
\delta_\bfX d\ell= d\ell \, (\dot{\Psi}_\parallel - \kappa_g \Psi_\perp  + \kappa_n \Phi) \,,
\end{equation}
proportional to the lhs of Eq.(\ref{delgproj}a). If this deformation is an isometry, then $\dot{\Psi}_\parallel = \kappa_g \Psi_\perp  - \kappa_n \Phi$. The first term on the right in Eq. (\ref{delS}), quantifying the response to a deformation along the tangent direction, is identified with the effect of a reparametrization of the curve: $\delta_\parallel d\ell = d \ell \,\dot{\Psi}_\parallel$. A scalar function $F(\ell)$, defined along the boundary curve, responds to a tangential deformation along the curve by $\delta_\parallel F = \dot{F} \,  \Psi_\parallel$; we thus see that $\delta_\parallel \int d\ell F = \int d\ell (F \Psi_\parallel)\, \dot{}$, so that the tangential deformation of the integrated $F$ vanishes if the curve is closed. 
\vskip1pc \noindent
Using Eq.(\ref{delS}) and integrating by parts, we have that the variation of $H_1$ is
\begin{equation} \label{boundaryarc}
\delta H_1 = \int d\ell\,\left[- \dot{\cal T} \, \Psi_\parallel + {\cal T} (- \kappa_g \Psi_\perp  + \kappa_n \Phi) \right]\,.
\end{equation}
The geodesic curvature is also an isometry invariant of the boundary curve. This invariance can also be imposed explicitly by adding to $H_C$ an addition constraint
\begin{equation}
H_2 = \int dv \, \Lambda (v) ( \sqrt{G} \kappa_g - \sqrt{G^{(0)}} \kappa_g^{(0)}) \,,
\end{equation}
involving a second local multiplier field $\Lambda$. Here $\kappa_g^{(0)}$ is the fixed geodesic curvature.
\vskip1pc \noindent
Now let us look at the variation of the geodesic curvature. A straightforward if long calculation, reproduced in \ref{App:derkappag}, gives the result
\begin{equation} \label{delkg}
\delta_\bfX \kappa_g = \dot{\kappa}_g \Psi_\parallel +  \ddot{\Psi}_\perp + (\kappa_g^2 + K_G ) \Psi_\perp - \kappa_g  \kappa_n \Phi - (\tau_g \Phi)\,\dot{} - \tau_g \dot{\Phi} - \kappa_n \nabla_\perp \Phi \,.
\end{equation}
When $\delta \bfX$ is an isometry, $\delta_\bfX \kappa_g = 0$.\footnote{Note that variation of the geodesic curvature in the orthogonal direction $\delta \kappa_{\perp \,g}$, involves $\nabla_\perp \Psi_\parallel$ as well as $\nabla_\perp \Phi$.} The terms involving $\Psi_\perp$ describe the deformation of the curvature as the curve is deformed along the surface. Significantly, the last term involves $\nabla_\perp \Phi$ (the problematic variation in Eq.(\ref{boundaryvar}), notably absent in the projections of $\delta_\bfX g_{ab}$ given by Eq.(\ref{delgproj}), or in Eq. (\ref{delS}), and thus \textit{not} implied by these constraints.
\vskip1pc \noindent
Combining Eq.(\ref{delkg}) with the variation of $\sqrt{G}$ given above Eq.(\ref{delS}), we have
\begin{equation} \label{delsqrtGkg}
\delta_\bfX \left( \sqrt{G} \, \kappa_g \right) = \sqrt{G} \left( (\kappa_g \Psi_\parallel) \, \dot{} +  \ddot{\Psi}_\perp + K_G \Psi_\perp - (\tau_g \Phi)\,\dot{} - \tau_g \dot{\Phi} - \kappa_n \nabla_\perp \Phi \right)\,.
\end{equation}
Therefore, up to a total derivative, the variation of $H_2$ is
\begin{eqnarray} \label{boundarygeod} 
\delta H_2 &=& \int d\ell\,\left[ - \dot{\Lambda} \, \kappa_g \, \Psi_\parallel + \Big(\ddot{\Lambda}  + \Lambda \, K_G  \Big) \,\Psi_\perp    + \Big(\dot{\Lambda}  \tau_g + (\Lambda  \tau_g )\,\dot{} \Big) \Phi  - \Lambda  \kappa_n \nabla_\perp \Phi\, \right]\,.
\end{eqnarray}
The constraints imposed by $H_1$ and $H_2$ are mutually independent. We claim that $H_1$ and $H_2$ together are not only necessary but also sufficient to ensure consistent boundary conditions on the constrained EL equations. 
\vskip1pc \noindent
Note that there is a fundamental difference between the constraints on the boundary and those in the interior. This is because the derivatives of bulk variations are surface covariant derivatives and we can peel off all such derivatives by integrating by parts as often as is necessary. 
With bulk multipliers imposing the constraint on the metric in place, the two tangential deformations as well as the normal deformation can be assigned independently. 
We do not possess this liberty on the boundary.  The boundary constraints reflect this  inability to vary the deformations and their derivatives  independently there.
As we will see the two constraints on the boundary play distinct roles.
\vskip1pc \noindent
In the context of $H_2$, it is relevant to note that the geodesic curvature of the boundary appears explicitly in the Gauss-Bonnet invariant,  
\begin{equation}
\int dA \,  K_G + \int d\ell \, \kappa_g \,.
\end{equation}
As we will see, one of its consequence is the redundancy of the boundary terms originating in the Gaussian curvature. 
\vskip1pc \noindent
How do we know that we possess a sufficient set of constraints on the boundary variations? One could imagine constraining isometry invariants involving higher derivatives along the boundary; for example, the Gaussian curvature, $K_G$, involving two derivatives. What we would find, however, is that the corresponding variation involves a term proportional to $\nabla^2_\perp \Phi$, a term that does not appear anywhere else in the variation of $H_C + H_1 + H_2$. Consistency would require treating this variation as independent which would imply that the corresponding multiplier must vanish, indicating the redundancy of this  constraint.  It would, however, be relevant in a theory including in the local bulk energy a term proportional to $(\nabla K)^2$. There is an important message here: the relevant local constraints on the boundary capturing isometry depend themselves on the order of the energy being considered.\footnote{Had one imposed a constraint on $K_G$ in the bulk, or indeed added additional constraints on tensors constructed with the metric, this would simply have amounted 
to a redefinition of  the multiplier $T^{ab}$ appearing in Eq. (\ref{VCdef}),
as one would have expected: there is no bulk consequence. However, boundary terms will arise in the process of \textit{peeling derivatives} off variations; the corresponding multipliers imposing these constraints---while redundant in the bulk---now imply boundary constraints: an interesting exercise/challenge is to identify the higher order bulk constraint induced boundary variations equivalent to those introduced by the sum, $H_1 + H_2$.}


\subsection{The correct boundary conditions}

By imposing the arc-length and geodesic curvature constraints, captured by Eqs. (\ref{delS}) and (\ref{delkg}) in the variational principle, we are free to  treat each of the four variations $\Psi_\parallel$, $\Psi_\perp$, $\Phi$ and $\nabla_\perp \Phi$ appearing in Eq. (\ref{boundaryvar}) as independent. To the expression for $\delta H_{C\,{\sf boundary}}$, given by Eq. (\ref{boundaryvar}), we thus add the sum of $\delta H_1$ and $\delta H_2$,  given respectively by Eqs. (\ref{boundaryarc}) and (\ref{boundarygeod}), obtaining 
\begin{eqnarray}
\delta H_{C\,{\sf boundary}} + \delta H_1 +\delta H_2 
&=& - \int d\ell\,  \left[ K + (\Lambda + \mathrm{k}_G ) \, \kappa_n \right] \, \nabla_\perp \Phi \\
&& + \int d\ell \, 
\left[\nabla_\perp K + \kappa_n  {\cal T} + 2 \dot{\Lambda} \tau_g  + (\Lambda + \mathrm{k}_G) \dot{\tau}_g\right] \, \Phi \nonumber \\
&& - \int d\ell\,
\left[T_{\parallel \perp}^{\calD} + \dot{\cal T} + \dot{\Lambda} \, \kappa_g \right]\,\Psi_\parallel \nonumber\\
&&- \int d\ell \, 
\left[ T_{\perp \perp}^{\calD}-\frac{1}{2} K^2 +  \kappa_g  {\cal T} - \ddot{\Lambda}  - (\Lambda + \mathrm{k}_G) K_G \right] \, \Psi_\perp \nonumber\,.
\end{eqnarray}
It is clear that, under the redefinition $\Lambda \to \Lambda + \mathrm{k}_G$, the Gaussian energy does not contribute to the boundary conditions. It would have been very strange if it did in view of its isometry invariance. An important corollary follows: an isotropic unstretchable sheet, flat or not, is characterized by a single bending modulus. 
\vskip1pc \noindent
In equilibrium, the four coefficients must vanish:  
\begin{subequations} \label{bcs}
\begin{eqnarray} 
K  +  \Lambda \, \kappa_n &=& 0 \,; \label{bc1}\\
\nabla_\perp K + \kappa_n {\cal T} + 2 \dot{\Lambda} \tau_g + \Lambda \dot{\tau}_g &=& 0 \,; \label{bc2}\\
T_{\parallel \perp}^{\calD} + \dot{\cal T}  + \dot{\Lambda} \kappa_g &=& 0\,; \label{bc3}\\
T_{\perp \perp}^{\calD} -\frac{1}{2} K^2   +  \kappa_g {\cal T} - \ddot{\Lambda} - \Lambda K_G &=& 0\,. \label{bc4}
\end{eqnarray}
\end{subequations}
While our specific interest is in flat surfaces, these boundary conditions are valid whether the surface is flat or not.
\vskip1pc \noindent
If the boundary is not asymptotic, so everywhere $\kappa_n\ne 0$,\footnote{The case $\kappa_n=0$ at a point must be treated separately. We will encounter an example in the context of cones. } then Eq. (\ref{bc1}) indicates that $\Lambda$ serves as a boundary catch-all for unbalanced torques on the boundary originating in the bulk, which it then feeds into all three remaining boundary conditions. So far so good.  Modulo this condition, Eq. (\ref{bc2}) reads
\begin{equation} \label{calT}
\nabla_\perp K + \kappa_n {\cal T} - 2  (K/\kappa_n)\,\dot{} \, \tau_g - K \dot{\tau}_g/\kappa_n =0\,,
\end{equation}
determining the second boundary multiplier ${\cal T}$ directly in terms of the boundary geometry. Importantly, neither of the two unknown as yet  isometric stresses, $T_{\| \perp}^{\calD}$ nor $T_{\perp \perp}^{\calD}$ appears in the first two boundary conditions (\ref{bcs}a) and (\ref{bcs}b) that have now been solved. But this should not be surprising: $T_{\| \perp}^{\calD}$ and $T_{\perp \perp}^{\calD}$ are tangential constraints passed from the bulk to the  boundary.
\vskip1pc \noindent
Along boundaries of a flat surface, coinciding with a non-trivial principal direction, the boundary normal curvature coincides with the mean curvature, $K=\kappa_n$, and Eq. (\ref{bcs}d) implies $\Lambda = -1$ (proportional to the rigidity modulus, which we have normalized); significantly, it does not vanish.\footnote{In this case the rulings are perpendicular to the boundary.}
Along a principal direction, $\tau_g = 0$ so that Eq. (\ref{calT}) also simplifies 
\begin{equation} \label{phi} \kappa_n {\cal T} = - \nabla_\perp \, \kappa_n\,, 
\end{equation}
fixing ${\cal T}$ in a very simple way in terms of the derivative of the normal curvature along the conormal, evaluated on the boundary.
\vskip1pc \noindent
Having determined $\Lambda$ and ${\cal T}$ and substituting the two into the remaining BCs (\ref{bc3}) and (\ref{bc4}), they determine the projections of the tangential stress, $T_{\| \perp}^{\calD}$ and  $T_{\perp \perp}^{\calD}$ at the boundary completely in terms of the boundary geometry. We now need to solve the shape equations subject to these boundary conditions. In general, the determination of the shape involves the boundary conditions. For cones, as we will see,  its shape is largely insensitive to the boundary conditions. However, to determine the stress that resides within the cone we are forced to confront these conditions. Let us first look at circular cones.


\section{Circular cones}

\begin{figure}
\begin{center}
  \includegraphics[height=5cm]{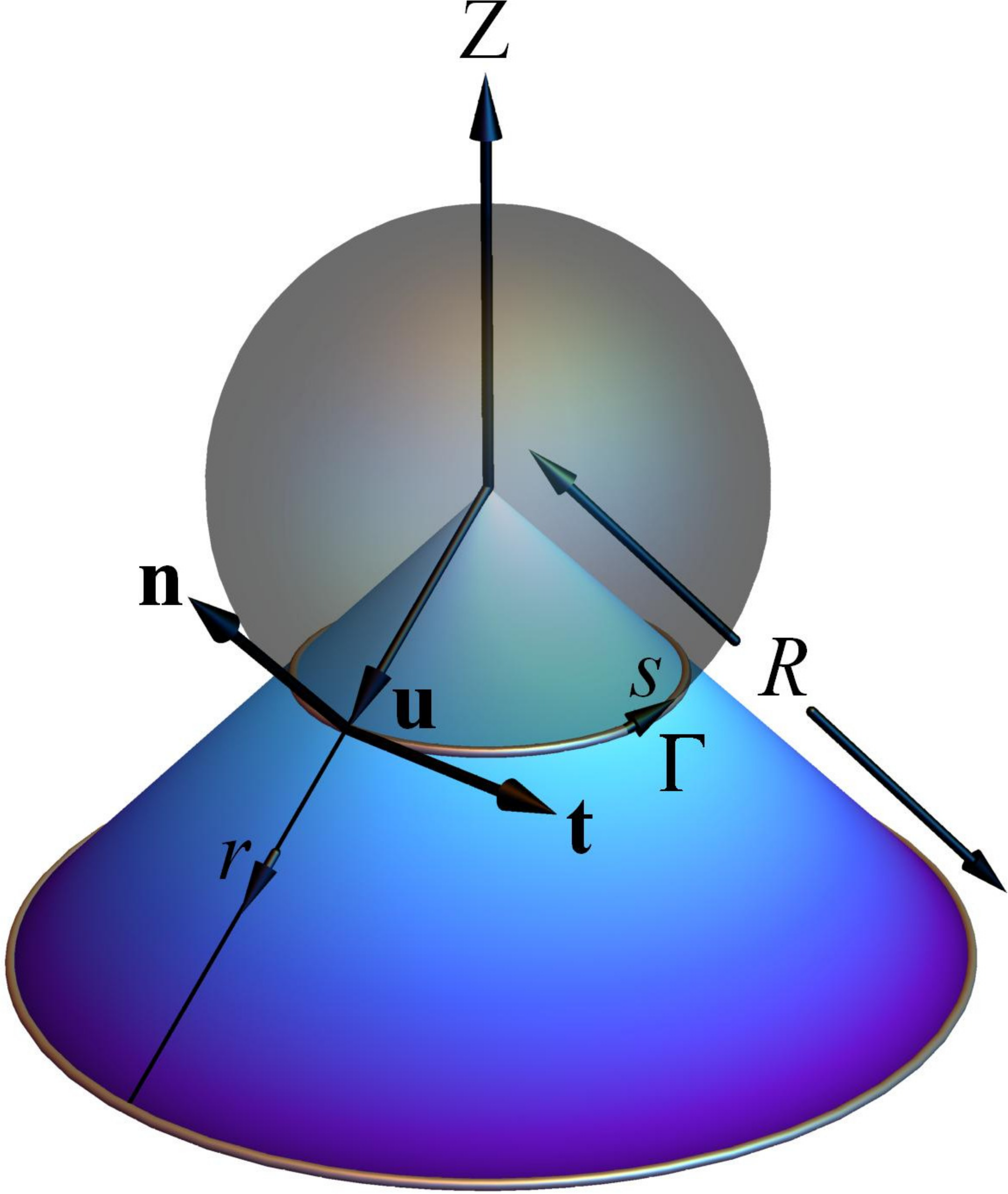}
\end{center}
\caption{Cone described by a curve $\Gamma$ on the unit sphere parametrized by arc length $s$.} \label{fig:regcone}
\end{figure}

A natural representation of a cone is provided by an arc-length parametrized closed curve $\Gamma: s\to {\bf u}(s)$ on the unit sphere $\mathbb{S}^2$. Let $r$ be the distance from the origin along the ray pointing in the direction ${\bf u}(s)$. The image of the mapping
\begin{equation}
(r,s) \mapsto  {\bf X}(r,s) = r {\bf u}(s)
\end{equation}
then describes a cone with its apex located at the origin, see Fig. \ref{fig:regcone}.
\vskip0pc \noindent
The tangent vectors to the cone adapted to this parametrization are ${\bf e}_r = \partial_r {\bf X} = {\bf u}$, and ${\bf e}_s= \partial_s{\bfX}= r {\bf u}'$, where the prime denotes derivatives with respect to $s$. Since ${\bf u}$ is parametrized by $s$, ${\bf t} ={\bf u}'$ is the unit tangent to $\Gamma$. Moreover, ${\bf u}$ is unit, so it is orthogonal to the tangent vector, ${\bf u}\cdot {\bf u}'=0$. Thus, the induced metric on the surface $g_{ab} = {\bf e}_a\cdot {\bf e}_b$ is given by the remarkably simple, manifestly flat, form 
\begin{equation} \label{eq:gab}
g_{ab}= 
\left(
\begin{array}{cc}
  1 & 0\\
  0 & r^2
\end{array}
\right) \,.
\end{equation}
The length of $\Gamma$ is given by 
\begin{equation} 
L = 2\pi + \Delta \varphi\,,
\end{equation}
where $\Delta \varphi$ is the angle difference at the apex. It is invariant under isometry. For a deficit angle $\Delta \varphi<0$, the closed curve representing such a cone lives always within a single hemisphere. $\Delta \varphi=0$ corresponds to a flat planar disc represented by a great circle. For surplus angle $\Delta \varphi>0$ the curve representing the cone necessarily spans both hemispheres.

\subsection{Cones as constrained equilibria} \label{Sec:coneconsteq}

Let ${\bf n} = {\bf u} \times {\bf t}$ denote the normal to the surface. The extrinsic curvature tensor, $K_{ab} = -{\bf n}\cdot \partial_a {\bf e}_b$ is given by
\begin{equation} \label{eq:Kab}
K_{ab} =  \left( 
\begin{array}{cc}
  0 & 0\\
  0 & \kappa \, r 
\end{array}
\right) \,,
\qquad
\mbox{where}
\qquad
\kappa = -{\bf t}' \cdot {\bf n} \,.
\end{equation}
The mean curvature of the cone is $K=\kappa/r$ and the Gaussian curvature vanishes $K_G=0$.
Consider now a cone bounded by a curve at a constant distance from the apex, say at $r=R$, as in Fig. \ref{fig:regcone}. The normal curvature along such a curve is given by $\kappa_n = \kappa/R$; the corresponding geodesic curvature is $\kappa_g = 1/R$, independent of $\Delta \phi$, or of $s$. Such a curve is a mapping of a circle of constant radius $R$, with a circular wedge of equal radius added or removed, so this independence is not surprising. It is simple to confirm that $\kappa$ is also the geodesic curvature along ${\bf u}(s)$ on the unit sphere. For a cone, the shape equation (\ref{EL:normal}) reduces to \cite{GM08}
\begin{equation}
\kappa'' + \frac{\kappa^3}{2} + \kappa  + \kappa r^2 T_{\parallel \parallel} =0\,,
\end{equation}
where $T_{\parallel \parallel} = \rmt^a \rmt^b K_{ab}$ is the projection of the constraining stress along $\mathbf{t}$. The first three terms are independent of $r$. Consistency completely determines the radial dependence of $T_{\parallel}$:
$T_{\parallel} =  - C_\parallel(s)/r^2$.\footnote{ Had we imposed the weaker constraint that the area is fixed, then $T^{ab}=-\sigma g^{ab}$, where $\sigma$ is a constant. In this case, a cone is never an equilibrium geometry.  Whereas isometry localizes the energy in the neighborhood of the apex,  under the weaker constraint the sheet is free to lower its energy by distributing the curvature away from the apex.} We now have  for the normal EL equation (\ref{EL}a), 
\begin{equation} \label{ELcone}
\kappa'' + \frac{\kappa^3}{2} + \left(1 - C_\parallel\right) \kappa =0\,,
\end{equation}
where $C_\parallel$ is generally a function of $s$. As shown in the appendix of \cite{GM08} and, as implemented in \cite{MBG08}, if the conical sheet is circular so that the boundary curve is a curve of  constant $R$, the rotational invariance of the energy implies that $C_\parallel$ is a constant in the absence of external forces. Equation (\ref{ELcone}) then reduces to a quadrature which can be solved using the boundary conditions associated with closure:
\begin{equation}
\kappa ((2\pi +\Delta\phi)/n) = \kappa(0) \,,
\qquad
\varphi(2\pi +\Delta\phi) = 2\pi \,,
\end{equation}
where $\varphi$ is the azimuthal angle on the spherical trajectory. Extraordinarily one does not need 
to solve the conservation law $\nabla_a T^{ab}=0$ to determine the equilibrium geometry.  To determine the stress distributed within the sheet, one not only needs to solve the conservation law, one also needs to possess the correct boundary conditions on the free edge of the cone. Had it not been for this \textit{happy} accidental uncoupling, when two of us first examined the problem back in 2008, we would have been obliged to address the inconsistency lurking in the boundary conditions on the free edge back then, not eleven years later.
\vskip1pc \noindent
If the sheet is not cut from a circle, or even if it is but it is manipulated  so as to break the rotational invariance, $C_\parallel$ will not be constant, the solution of the shape equation (\ref{ELcone})  needs to be supplemented with the boundary distance to the apex. 
This issue will be taken up in Sec. \ref{Cones with non-circular boundary}.
\vskip1pc \noindent
The bulk stress $T^{ab}$ associated with isometry can be determined from the shape equation 
Eq. (\ref{EL}a) and the conservation law Eq. (\ref{EL}b), and, as shown in reference \cite{GM08}, 
is given in terms of $C_\parallel$ and two additional unknown functions of $s$, $C_{\parallel\perp}(s)$ and $C_{\perp}(s)$ (recall $\bfl= - \mathbf{u}$) 
\begin{subequations} \label{def:Tparperp}
\begin{eqnarray}
T_{\parallel\parallel}(s,r) &:=& \rmt^a \rmt^a T_{ab} = - \frac{C_\parallel}{r^2}\,, \label{Tparpar}\\
T_{\parallel\perp}(s,r) &:=& \rml^a \rmt^b T_{ab} = - \frac{1}{r^2} \left( C_\parallel' \ln r + C_{\parallel\perp}\right)\,, \label{Tparperp}\\
T_{\perp\perp}(s,r) &:=& \rml^a \rml^b T_{ab}= \frac{1}{r^2} \left(C_\parallel'' (\ln r + 1) + C_\parallel + C_{\parallel\perp}'\right) + \frac{C_\perp}{r} \,. \label{Tperpperp}
\end{eqnarray}
\end{subequations}
Unless the boundary is a curve of constant $r$, the tangent and conormal to $\Gamma$ $\{\bft, \bfl\}$ will not be parallel to the tangent and conormal adapted to the boundary $\{\bfT,\bfL\}$, so these projections generally will not coincide with the projections $T_{\perp \|}^{\calD}$ and $T_{\perp \perp}^{\calD}$ appearing in the boundary conditions in Sec. (\ref{Sec:FBC}).
\\
For a cone with a fixed outer radius $r=R$, the boundary forms a principal direction, $K=\kappa_n=\kappa/R$, and $\tau_g=0$. As a result of the former, Eq.(\ref{bc1}) implies $\Lambda =-1$. Equation (\ref{bc2}), (or Eq.(\ref{phi})), in turn, implies that ${\cal T}=-1/R$. For these values of $\Lambda$ and ${\cal T}$, the boundary condition (\ref{bc3}) now implies that $T_{\parallel\perp}^{\calD}=0$. 
Moreover, $T_{\parallel \perp} (s,R) = T_{\parallel \perp}^{\calD}$, and from Eq. (\ref{Tparperp}) follows that there will be a shearing stress induced by the spatial variation of $C_{\parallel}$
\begin{equation} \label{eq:Cparperpcone}
C_{\parallel \perp}(s) = - C_\parallel' \ln R \,. 
\end{equation}
Thus, $C_{\parallel \perp}$ vanishes when $C_{\parallel}$ is constant.
\vskip1pc \noindent
There remains to satisfy (\ref{bc4}) which implies that
\begin{equation}
T_{\perp\perp}^{\calD}(s) = \frac{1}{R^2} \left(\frac{\kappa^2}{2} + 1 \right)\,.
\end{equation}
Since, $T_{\perp \perp} (s,R) = T_{\perp \perp}^{\calD}(s)$, Eq. (\ref{Tperpperp}) determines $C_\perp(s)$:  
\begin{equation} \label{eq:Cperpcone}
C_\perp (s) = \frac{1}{R} \left(\frac{\kappa^2}{2} + 1 - C''_\parallel - C_\parallel \right)\,.
\end{equation}
Thus, within the bulk,  $T_{\parallel \perp}$ and $T_{\perp\perp}$ are completely determined by $C_\parallel$ and $\kappa$:
\begin{subequations} \label{Tprojdef}
\begin{eqnarray}
T_{\parallel \perp} (r,s)& = & -\frac{C'_\parallel}{r^2}\ln \frac{r}{R}\,;\\
T_{\perp\perp} (r,s) & = & \frac{C''_\parallel}{r^2} \left(\ln \frac{r}{R}+1-\frac{r}{R}\right) + \frac{C_\parallel}{r^2}\left(1-\frac{r}{R}\right)+\frac{1}{r R} \left(\frac{\kappa^2}{2} + 1 \right)\,.
\end{eqnarray}
\end{subequations}
Notice that, even when $C_\parallel$ is constant, $T_{\perp\perp}$ has a non-trivial curvature dependence. Notice also that, as the boundary $R\to \infty$, the stress vanishes as $r \to \infty$. 
\vskip1pc \noindent
To identify the complete stress within the bulk, we need to add the bending tangential stresses given by  Eq. (\ref{fB}):
\begin{subequations} \label{fBproj}
\begin{eqnarray}
f_{B\,\parallel \parallel} & = & \frac{\kappa^2}{2\,r^2}\,;\\
f_{B\,\parallel \perp} & = & 0\,;\\
f_{B\,\perp \perp} & = & - \frac{\kappa^2}{2 r^2}\,.
\end{eqnarray}
\end{subequations}
Notice that $f_{B}^{ab}$ is diagonal with respect to the principal vector fields. This is a consequence of the fact that $f_{B}^{ab}$ is a polynomial in the extrinsic curvature. More significantly, if the stress is compressive in one direction, it must be equally tensile in the orthogonal direction: $f_{B\,\perp\perp} = - f_{B\,\parallel\parallel}$. The latter is a direct consequence of the scale invariance of the two-dimensional bending energy \cite{CG02}. 

\subsection{The stressed paper cone} \label{stresscone}

In a deficit cone cut from a circular sheet, $\kappa = -\cot\theta_0$ is constant, where $\theta_0$ is the opening angle with respect to the axis of the cone, which is related to the deficit $\Delta \varphi$ by $\sin \theta_0 = 1 - |\Delta \varphi|/(2\pi)$. In this case, if $\kappa \ne 0$, the shape equation (\ref{ELcone}) implies that $C_\parallel = \kappa^2/2 + 1$, with $T_{\parallel\parallel}= - C_\parallel/r^2 =- T_{\perp\perp}$. This implies that the tangential stress associated with isometry is non vanishing for any $\Delta\varphi <0$, no matter how small!  As $\Delta \varphi \to 0$, $C_\parallel\to 1$! (reminiscent of an Euler buckling instability); as $\Delta \varphi\to - 2\pi$, $\sin \theta_0  \rightarrow 0$, so $\kappa$ and $C_\parallel$ diverge. If we add the contributions to the bending stress, given by Eqs. (\ref{fBproj}), we have\footnote{To determine forces, we should rightly add the contribution along $\mathbf{n}$, but these vanish in a circular cone.}
\begin{subequations}
\begin{eqnarray}
f_{\parallel\parallel} &= & -\frac{1}{r^2}\,;\\
f_{\parallel\perp}  &= & 0\,;\\
f_{\perp\perp} &= & \frac{1}{r^2} \,.
\end{eqnarray}
\end{subequations}
Thus the cone is under tension along circles of constant $r$, the qualitative behavior one would have anticipated. But now it has been quantified. Cut the cone along a ruling and it will splay open (symmetrically) into a planar circular wedge. Less obvious, it is under equal compression along its rulings. The attentive reader will also notice that the total stress---summing the isometric constraining and bending stresses---is independent of $\Delta \varphi$, the result of a cancellation occurring when they are added. This is a non-trivial prediction. Simple as this example may be, the distribution of stress could not have been determined without having access to the  correct boundary conditions on the rim. 
\\\\
In \ref{Force Balance}, we show explicitly that the correct balance of forces across the boundary involve the local boundary constraints in an essential way. 

\subsection{Turning deficit to surplus}

Let us briefly comment on the case where there is a surplus angle at the apex, with $\Delta \varphi>0$. The equilibrium shapes, examined in their full glory in \cite{MBG08}, exhibit rather non-trivial behavior.  In the spirit of this paper, however, we will look only at small surpluses here. If $\kappa \ll 1$, Eq. (\ref{ELcone}) linearizes, with solutions $\kappa=k_0\cos (ns)$, $n=2,3,4,\dots$, representing regular conical ripples about a planar annulus.  The constant $C_\parallel$ determining the constraining stress across the cone is  given by $C_\parallel = 1- n^2 < 0$. Note that $C_\parallel$ does not vanish in the limit $\Delta\varphi\to 0$ ($k_0\to 0$), a result that is analogous to an Euler-buckling instability. In this regime
\begin{subequations}\label{Tprojsurp}
\begin{eqnarray}
T_{\parallel\parallel} (r) &=& \frac{n^2 -1}{r^2}\,; \\
T_{\parallel \perp} & = & 0\,; \\
T_{\perp\perp} (r) & = & \frac{1-n^2}{r^2} \left(1-\frac{r}{R}\right)+\frac{1}{r R} \,,
\end{eqnarray}
\end{subequations}
whereas $f_B^{ab} \approx  0$. In this limit, what stress exists in the cone is due to the isometry constraint. 
This time $f_{\parallel\parallel}$ is compressive  everywhere. Cut the cone along a ruling and the sheet will relax into a  
planar circular wedge of angular width $2\pi +\Delta\varphi$.\footnote{If $\Delta \varphi$ is increased,  the stress will exhibit a dependence on $s$. Above some finite surplus, the geometry will develop overhangs; within these  overhangs $f_{\parallel\parallel}$ will become tensile.} 
The radial stress has non-trivial behavior: close to the apex, $f_{\perp\perp}\approx T_{\perp\perp}$  is tensile but beyond a critical radius, $r_c/R= 1 - 1/n^2$, it turns compressive. 

\section{Cones with non-circular boundary} \label{Cones with non-circular boundary}

\begin{figure}
\begin{center}
  \includegraphics[height=5cm]{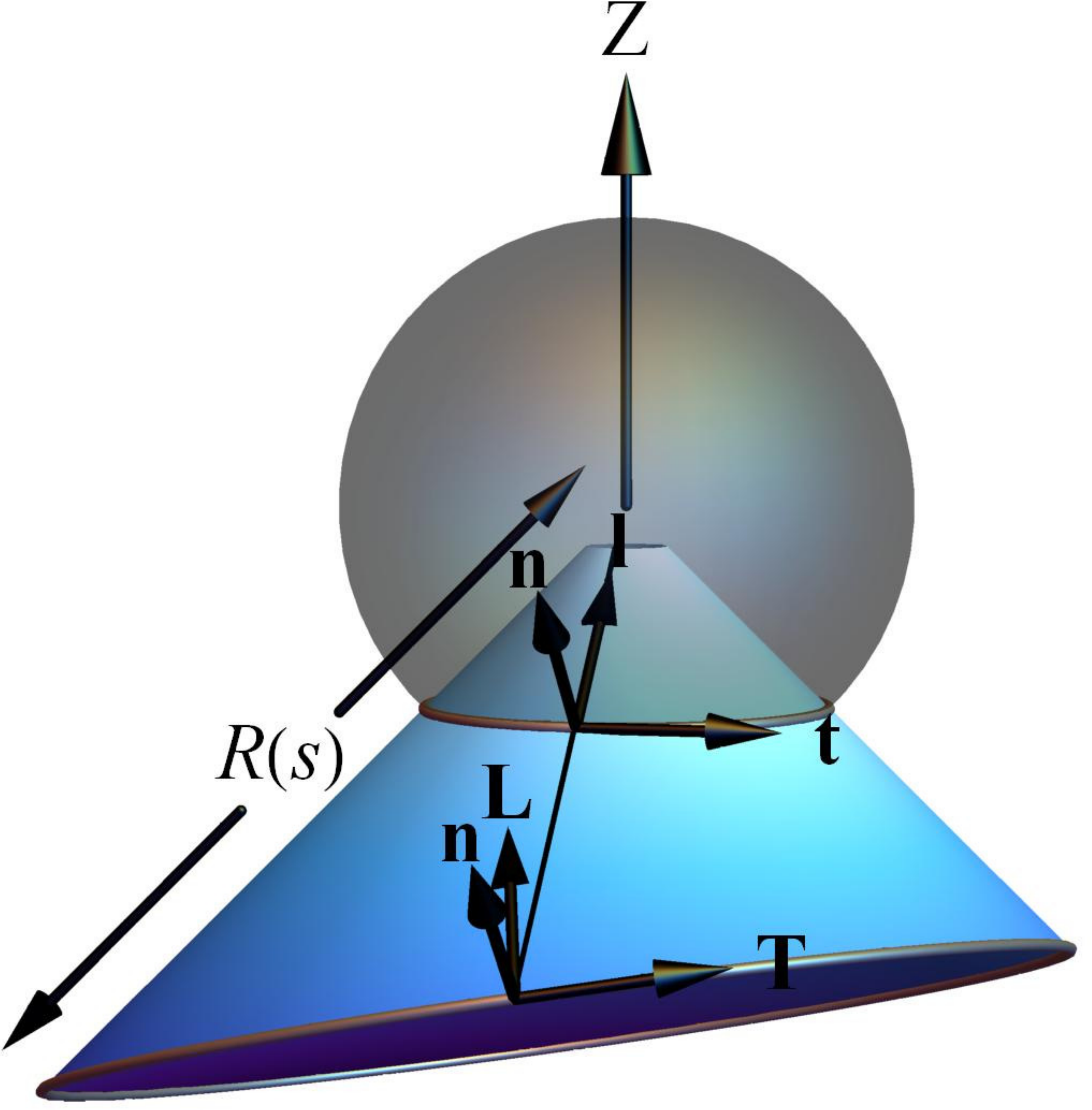}
\end{center}
\caption{Cone with a tilted boundary. The adapted Darboux frames to the spherical curve and the boundary 
are given by the trihedrons $({\bf t}, {\bf l} = -{\bf u}, {\bf n})$, and $({\bf T}, {\bf L}, {\bf n})$.} \label{fig:conetiltedbound}
\end{figure}

So far we have focused on cones isometric to a segment of a circular planar sheet. 
Let us now consider a deficit cone with an arbitrary boundary parametrized by the arc length of ${\bf u}$, $R = R(s)$  such as that illustrated in Fig \ref{fig:conetiltedbound}. We will show that the corresponding equilibrium cone responding to this boundary is no longer circular.
\footnote{For simplicity, the inner boundary remains \textit{circular}, in that  the cutoff distance from the apex,  $r_0$, introduced to regularize the energy, is independent of $s$.}\label{f6}  In this case, since $\kappa = {\bf u}'' \cdot {\bf u}' \times {\bf u} $, the bending energy $H_B$ is a functional not only of the curve ${\bf u}$, but also of the boundary 
\begin{equation}
\label{eq:HB}
H_B[{\bf u}, B] = \frac{1}{2} \oint ds\, B(s) \, \kappa^2 \,, \quad B(s) = \ln \left( \frac{R(s)}{r_0}\right)\,,
\end{equation}
where $r_0$ is a cutoff distance from the apex mentioned in footnote 12. The energy is the Euler elastic energy on a sphere with a bending rigidity $B$ that depends explicitly on the position along the curve (recall it is dimensionless because the bending rigidity itself is).
The breaking of rotational symmetry here
gives rise to a source in the Euler-Lagrange equation for $\kappa$, analogous to a \textit{material force}, as discussed recently by O’Reilly \cite{O'Reilly2017}.
Even in the absence of external forces, if $B$ is not constant, the geometry is not generated by a circle of  constant $\kappa$.  We can see this by solving the EL equation  for the energy (\ref{eq:HB}) with the boundary described by the distance function $R(s)$ (and the length $L$ it implies) to determine $\kappa$ for the equilibrium geometry.  

\subsection{Determination of $\kappa$ and $C_{\parallel}$}

Let us first derive  the EL equations that correspond to $H_B$,  given by Eq. (\ref{eq:HB}).  If $R$ is not constant, it is evident that $\kappa$  will depend explicitly on $R$ through the function $B(s)$. To do this we need to track the change in $H_B$ under a deformation of ${\bf u}$, holding the arc-length fixed.\footnote{A derivation for constant $B$ was presented in the appendix of Ref. \cite{GM08}.}  With respect to the adapted basis, a deformation of the unit vector ${\bf u}$ reads $\delta {\bf u} = \delta \psi  \, {\bf t} + \delta \phi \, {\bf n}$. Using the commutation of the variation and arc-length derivative we determine the variation of the tangent vector
\begin{equation} \label{eq:deftn}
\delta {\bf t} = (\delta {\bf u})' = -\delta \psi {\bf u} + (\delta \phi' - \kappa \delta \psi) {\bf n}\,.
\end{equation}
The fixed length constraint ${\bf t} \cdot \delta {\bf t}=0$ places the following constraint on the components of $\delta {\bf u}$
\begin{equation} \label{eq:isomcond}
\delta \psi' + \kappa \delta \phi=0\,.
\end{equation}
The deformed normal vector $\delta {\bf n}$ follows from the orthonormality of the basis. Combining these expressions we identify the variation of $\kappa$, 
\begin{equation}
 \delta \kappa = -\delta \phi'' -\delta \phi  + (\kappa \delta \psi)' \,.
\end{equation}
Substituting this result into the variation of $H_B$, $\delta H_B = \int ds B \kappa \delta \kappa$, and integrating by parts,\footnote{We neglect the total derivatives because ${\bf u}$ is a closed curve.} we find that 
\begin{equation} \label{eq:deltaHB1}
\delta H_{B} = - \oint ds \left[ \frac{B'}{2} \kappa^2  \delta \psi + \left( (B \kappa)'' + B \kappa \left( \frac{\kappa^2}{2} + 1\right) \right) \delta \phi \right] \,.
\end{equation}
We see that the arc length dependence of $B$ prevents us from discarding the tangential component of the deformation in the variation of the energy. The way we proceed is to use the isometry constraint (\ref{eq:isomcond}) to replace $\delta \phi$ in favor of $\delta \psi'$ (assuming $\kappa \neq 0$ everywhere).  This approach turns conventional wisdom on its head (we tend to think of the tangential deformation in an isometry as a response to the normal deformation), but it is completely legitimate no matter how strange it may appear.  It has been used recently in a related context \cite{Arrest}.  The price paid is a higher order EL equation. On conducting a final  integration by parts we obtain
\begin{equation} \label{eq:deltaHB2}
\delta H_{B} = -\oint ds \left[ \left( \frac{(B \kappa)''}{\kappa} + B \left( \frac{\kappa^2}{2} + 1\right) \right)' + \frac{B'}{2} \kappa^2 \right] \delta \psi \,.
\end{equation}
Therefore, in equilibrium we have the EL equation
\begin{equation} \label{eq:ELeq3}
 \left( \frac{(B \kappa)''}{\kappa} + B \left( \frac{\kappa^2}{2} + 1\right) \right)' + \frac{B'}{2} \kappa^2 = 0 \,.
\end{equation}
Integration of this equation along $\Gamma$ places an integrability condition on $B$ and $\kappa$
\begin{equation} \label{eq:intcondBkappa}
\oint ds B' \kappa^2=0\,.
\end{equation}
Furthermore, we have the first integral
\begin{equation} \label{eq:akappapp}
(B \kappa)'' + \kappa \, \left(B \left(\frac{\kappa^2}{2} + 1\right) +\frac{1}{2} \int_0^s ds \, B' \kappa^2 - c \right) = 0 \,,
\end{equation}
where $c$ is a constant of integration. As we will see explicitly, in our perturbative treatment,  $c$ is determined by the requirement of closure.  
Consistency of the first integral (\ref{eq:akappapp}) with the bulk EL Eq.(\ref{ELcone}) determines $C_{\parallel}$ as a function of $s$ through $B(s)$ and the yet to be determined $\kappa$:
\begin{equation} \label{eq:Cparper}
C_{\parallel}(s) = -\frac{1}{B} \left(B'' + 2 B' \frac{\kappa'}{\kappa} + \frac{1}{2} \int_0^s ds \, B' \kappa^2 -c \right) \,.
\end{equation}
If the boundary is circular with constant $R$ (or $B$), we have $C_\parallel = c/B$, a constant. 
In general, $C_\parallel$ is not  constant.  Now let us examine the construction of the curve and how the constant $c$ is determined. 

\subsection{Reconstruction of the curve}

To determine the embedding functions of the non-circular boundary, we parametrize ${\bf u}$ by spherical coordinates dependent on $s$
\begin{equation}
{\bf u}(s) = (\sin \theta(s) \cos \phi(s), \sin \theta(s) \sin \phi(s), \cos \theta(s))\,.
\end{equation}
The basis vectors adapted to the curve (${\bf u} = \hat{\bf r}$) can be expressed
\begin{subequations} \label{def:tnsphere}
 \begin{eqnarray}
{\bf t} &=& \theta' \,\hat{\bm \theta} + \sin \theta \phi' \, \hat{\bm \phi} \,,\\
{\bf n} &=& - \sin \theta \phi'  \, \hat{\bm \theta} + \theta' \, \hat{\bm \phi} \,,
\end{eqnarray}
\end{subequations}
in terms of the spherical basis\footnote{$\theta$ is measured from the positive $Z$ axis, whereas $\phi$ in the anticlockwise sense about the $Z$ axis.}
\begin{subequations}
 \begin{eqnarray}
 \hat{\bm \theta} &=& (\cos \theta(s) \cos \phi(s), \cos \theta(s) \sin \phi(s), -\sin \theta(s)) \,, \\
\hat{\bm \phi} &=& (- \sin \phi (s), \cos \phi (s),0) \,,
\end{eqnarray}
\end{subequations}
Notice that $\bf n = {\bf u} \times {\bf t}$ is the outward normal to the cone. The normalization of the basis requires the following relation between the derivatives of the two spherical angles
\begin{equation} \label{normrel}
 \left(\theta '\right)^2+\sin ^2 \theta \left(\phi '\right)^2 = 1\,.
\end{equation}
Thus $\phi'$ is given by\footnote{We consider only the positive root, for the negative one just represents the same curve but traversed in the opposite sense.}
\begin{equation} \label{phip}
 \phi ' = \csc \theta \sqrt{1-(\theta')^2}\,.
\end{equation}
Differentiating this relation we obtain
\begin{equation} \label{phipp}
 \phi '' = -\frac{\csc \theta}{\sqrt{1-\theta'{}^2}} \left(\theta'' + \cot \theta (1-\theta'{}^2) \right) \,.
\end{equation}
Now, we can calculate $\kappa = -{\bf t}' \cdot {\bf n}$: taking the arc length derivative of the first of Eqs. (\ref{def:tnsphere})\footnote{The derivatives of the spherical tangent basis are
\begin{subequations}
 \begin{eqnarray}
\hat{\bm \theta}' &=& - \theta' \hat{\bf u} + \cos \theta \phi' \hat{\bm \phi}  \,,\\
 \hat{\bm \phi}' &=& -\phi' ( \sin \theta \hat{\bf u} + \cos \theta\, \hat{\bm \theta}) \,.
\end{eqnarray}
\end{subequations}
} projecting it onto the expression of ${\bf n}$ and substituting the expressions (\ref{phip}) and (\ref{phipp}) for the derivatives of $\phi$, we obtain
\begin{equation}
\kappa =  \frac{\theta''-\cot \theta \left(1-\theta'{}^2\right)}{\sqrt{1-\theta'{}^2}} \,.
\end{equation}
We can recast this equation as a second order differential equation for $\theta$,
\begin{equation} \label{DEtheta}
 \theta '' - \cot \theta \left( 1-\left(\theta '\right)^2 \right) - \kappa \sqrt{1-\left(\theta ' \right)^2} = 0 \,. 
\end{equation}
One needs to solve Eqs. (\ref{eq:ELeq3}), (\ref{DEtheta}) simultaneously; the constant $c$ is then determined by integrating Eq. (\ref{phip}) for $\phi(s)$ and demanding closure. Having determined $\kappa$ and $c$, it is straightforward to use the identity (\ref{eq:Cparper}) to evaluate $C_\|(s)$, which completes the determination of $f_{\|\|}(r,s)$. In full generality this is a non-trivial numerical exercise. In perturbation theory, however, it becomes tractable. 


\subsection{Determination of the boundary stresses}
\label{bstresses}

We parametrize the boundary of the cone by $R(s)$: ${\bf X}_b = R(s) {\bf u}(s)$. The tangential vectors of the Darboux frame adapted to the boundary, $\{\bfT, \bfL \}$, are related to the tangential vectors of the Darboux frame of the spherical curve $\bf{u}$, $\{\bf{t}, \bf{l} \}$\footnote{The unit normal vector is ${\bf n}$ in both frames, because it can be parallel transported along rulings.} (see Fig. \ref{fig:conetiltedbound}) by a rotation
\begin{subequations} \label{eq:TLspnbytl}
\begin{eqnarray}
{\bf T} &=& \frac{1}{\sqrt{R'{}^2 + R^2}} (R \, {\bf t} - R' \, {\bf l}) \,,\\
{\bf L} &=& \frac{1}{\sqrt{R'{}^2 + R^2}} (R' \, {\bf t} + R \, {\bf l}) \,.
\end{eqnarray}
\end{subequations}
The boundary line element is $d \ell = \sqrt{R'{}^2 + R^2} ds$, so arc-length derivatives are related by $\partial_\ell = \frac{1}{\sqrt{R'{}^2 + R^2}} \partial_s$. Substituting these relations into Eqs. (\ref{BndyDbxeqs}) for the Darboux curvatures of the boundary, we determine 
\begin{subequations}
\begin{eqnarray} \label{eq:Dbxcrvscnbdr}
\kappa_{g} &=& \frac{1}{\sqrt{R'{}^2 + R^2}} \left(2 - \frac{R (R'' + R)}{R'{}^2 + R^2} \right)\,, \\
\kappa_{n} &=& \frac{\kappa R}{R'{}^2 + R^2}\,, \\
\tau_{g} &=& -\frac{\kappa R'}{R'{}^2 + R^2} \,,
\end{eqnarray}
\end{subequations}
along the boundary.
The normal curvature in the direction orthogonal to the boundary is determined from the planarity condition, $K_G = \kappa_n \kappa_{n \perp} -\tau^2_g =0$, which yields
\begin{equation}
\kappa_{n \perp} = \frac{\kappa R'{}^2}{R \left(R'{}^2 + R^2\right)}\,.
\end{equation}
Thus $K = \kappa_n + \kappa_{n \perp} = \kappa/R$, depending only passively on the boundary, and consistent with Eq.(\ref{eq:Kab}). Since $\bft= \bfe_s/R$ and $\bfl = -{\bf e}_r$,\footnote{The boundary is traversed in an anticlockwise sense with the bulk at the left-hand side, so $\bfL$ points towards the apex.} we have $\rmL^s =R'/(R \sqrt{R'{}^2+R^2})$ and $\rmL^r =-R/\sqrt{R'{}^2+R^2}$, so (using $K=\kappa/r$ in the bulk)
\begin{equation}
\nabla_\perp K = \frac{1}{R \sqrt{R'{}^2+R^2}} \left( \frac{R'}{R}\kappa' +\kappa \right)\,.
\end{equation}
From Eqs. (\ref{bc1}) and (\ref{bc2}) we determine $\Lambda$ and $\calT$ as functions of $\kappa$, $R$ and their derivatives
\begin{subequations}
\begin{eqnarray} \label{eq:LambdacalTbdrcn}
\Lambda &=& 
- \left(\frac{R'}{R}\right)^2 - 1\,; \\
\calT &=& -\frac{R}{\sqrt{R'^2 + R^2}} \, \left(\frac{R'}{R^2} \left(\frac{R'^2}{R^2}+1\right) \right)' - \frac{\sqrt{R'^2 + R^2}}{R^2} \left( 2 \frac{R' \kappa'}{R \kappa} + 1 \right)\,.
\end{eqnarray}
\end{subequations}
Eqs. (\ref{bc3}) and (\ref{bc4}) now determine $T_{\parallel \perp}$ and $T_{\perp \perp}$ at the boundary
\begin{subequations} \label{TparprpTprpbvar}
\begin{eqnarray}
T_{\parallel \perp}^{\calD}(s) &=& \frac{1}{\sqrt{R'^2 + R^2}}\left[\frac{R}{\sqrt{R'^2 + R^2}} \, \left(\frac{R'}{R^2} \left(\frac{R'^2}{R^2}+1\right) \right)' + \frac{\sqrt{R'^2 + R^2}}{R^2} \left( 2 \frac{R' \kappa'}{R \kappa} + 1 \right) \right]' \nonumber \\
&&+\frac{1}{R'^2 + R^2}\, \left(2-\frac{R \left(R'' + R \right)}{R'^2 + R^2}\right) \, \left(\frac{{R'}^2}{R^2}\right)' \,; \\
T_{\perp \perp}^{\calD}(s) &=&  \left(2-\frac{R \left(R'' + R\right)}{{R'}^2 + R^2}\right) \, \left[\frac{R}{{R'}^2 + R^2} \, \left(\frac{R'}{R^2} \,  \left(\frac{{R'}^2}{R^2}+1\right)\right)' + \frac{1}{R^2} \left( 2 \frac{R' \kappa'}{R \kappa} +1 \right)\right] \nonumber \\
 && - \frac{1}{\sqrt{{R'}^2+R^2}} \, \left[\frac{1}{\sqrt{{R'}^2+R^2}} \, \left(\frac{{R'}^2}{R^2}\right)' \, \right]' +\frac{\kappa^2}{2 R^2} \,.
\end{eqnarray}
\end{subequations}
The boundary tangential stresses are related to their bulk counterparts by a rotation. To this end, we invert Eqs. (\ref{eq:TLspnbytl}) to express the components of the tangent basis of the spherical curve in terms of those of the tangent basis adapted to the boundary as
\begin{equation} \label{eq:tlcompsTL}
\rmt^a = \frac{1}{\sqrt{R'{}^2 + R^2}} \left( R \rmT^a + R' \rmL^a\right) \,, \quad \rml^a = \frac{1}{\sqrt{R'{}^2 + R^2}} \left( -R' \rmT^a + R \rmL^a\right)\,.
\end{equation}
Substituting these expression into Eqs. (\ref{def:Tparperp}), we relate the bulk tangential stresses evaluated at the boundary in terms of the boundary stresses
\begin{subequations}
\begin{eqnarray} \label{eq:Tproys}
T_{\parallel \parallel} (s,R) &=&  \frac{1}{R'{}^2 + R^2} \left(R^2 T_{\parallel \|}^{\calD} + 2 R R' T_{\perp \|}^{\calD} + R'{}^2 T_{\perp \perp}^{\calD}\right)=-\frac{C_\parallel}{R^2}\,,\\
T_{\perp \parallel} (s,R) &=& \frac{1}{R'{}^2 + R^2} \left( \left( R^2 - R'{}^2 \right) T_{\perp \|}^{\calD}  + R R' \left( T_{\perp \perp}^{\calD} - T_{\parallel \|}^{\calD} \right)\right)= - \frac{1}{R^2} \left(C_\parallel' \ln R + C_{\parallel \perp} \right) \,, \\
T_{\perp \perp} (s,R) &=& \frac{1}{R'{}^2 + R^2} \left( R^2  T_{\perp \perp}^{\calD}  - 2 R R' T_{\perp \|}^{\calD} + R'{}^2  T_{\parallel \|}^{\calD} \right) = \frac{1}{R^2} \left(C_\parallel'' \left(\ln R + 1 \right) + C_\parallel  + C_{\perp \parallel}'\right) + \frac{C_\perp}{R}\,. \quad
\end{eqnarray}
\end{subequations}
We see that only when $R$ is constant we have $T_{\parallel \parallel} (s,R)= T_{\parallel \|}^{\calD}(s)$, $T_{\perp \parallel} (s,R)= T_{\perp \|}^{\calD}(s)$ and $T_{\perp \perp} (s,R)= T_{\perp \perp}^{\calD}(s)$. There are three unknowns $T_{\parallel \|}^{\calD}$, $C_{\parallel \perp}(s)$ and $C_{\perp}(s)$. 
From Eq. (\ref{eq:Tproys}), we determine the missing projection on the boundary, $T_{\parallel \|}^{\calD}$ modulo the functions $C_\|$, $R$ and $\kappa$:
\begin{equation} \label{eq:Tparparb}
T_{\parallel \|}^{\calD} = - \frac{C_{\parallel}}{R^2} \left(1 + \left(\frac{R'}{R}\right)^2 \right) - 2 \frac{R'}{R} T_{\perp \|}^{\calD} - \left(\frac{R'}{R} \right)^2 T_{\perp \perp}^{\calD}\,,
\end{equation}
where $T_{\perp \perp}^{\calD}$ and $T_{\perp\parallel }^{\calD}$  are given by Eqs. (\ref{TparprpTprpbvar}).
Substituting this result in  equations (\ref{eq:Tproys}b) and (\ref{eq:Tproys}c), determines the two remaining functions $C_{\parallel \perp}$ and $C_\perp$,
\begin{subequations} \label{CparprpCprpbvar}
 \begin{eqnarray}
C_{\parallel \perp}(s) &=& - R^2 T_{\parallel \perp}^{\calD}(s) - R R' T_{\perp \perp}^{\calD}(s) - \left(C_\parallel \, \ln R \right)' \,,\\
C_{\perp}(s) &=&  \left( R' \, T_{\perp \perp}^{\calD}(s) \right)' + R \left( T_{\parallel \perp}^{\calD}{}'(s) + T_{\perp \perp}^{\calD}(s) -\left( \frac{C_\parallel'}{R^2} \right)' \right)  - C_\parallel \left(\frac{1}{R} - \left(\frac{R'}{R^2}\right)' \right) \,.
\end{eqnarray}
\end{subequations}
Substituting these expressions back in Eqs. (\ref{Tparperp}) and (\ref{Tperpperp}), the tangential stresses are now completely determined in the bulk
\begin{subequations}
\begin{eqnarray}
T_{\parallel \perp}(r,s) &=& - \frac{1}{r^2} \left( C'_\parallel \, \ln \frac{r}{R} -C_\parallel \frac{R'}{R} -R^2 T_{\parallel \perp}^{\calD}(s) - R R' T_{\perp \perp}^{\calD}(s) \right) \,,\\
T_{\perp \perp}(r,s) &=& \frac{1}{r^2} \left[C_{\parallel}'' \left(1-\frac{r}{R}+\ln \frac{r}{R}\right) - 2 C_{\parallel}' \frac{R'}{R} \left(1-\frac{r}{R}\right) + C_{\parallel} \left( \left(1-\frac{r}{R}\right) \left(1 - \left(\frac{R'{}}{R} \right)' \right) - \frac{r}{R} \left( \frac{R'}{R}\right)^2\right)\right] \nonumber \\
&&-\frac{1}{r^2} \left(R R' T_{\perp \perp}^{\calD} + R^2 T_{\parallel \perp}^{\calD} \right)' + \frac{1}{r} \left( \left(R' T_{\perp \perp}^{\calD}\right)' + R T_{\perp \perp}^{\calD} + R T_{\parallel \perp}^{\calD}{}'\right) \,.
\end{eqnarray}
\end{subequations}
For constant $R$ these results reduce to those presented in Sec. \ref{Sec:coneconsteq}.
\\
As we can see, determining the cone and its stresses when the boundary is not circular is a highly coupled non-linear problem. We will confine ourselves to treating the problem perturbatively.

\subsection{Perturbative regime}
\label{section:perp}

\begin{figure}[htb]
\begin{center}
\subfigure[]{\includegraphics[width=0.32 \textwidth]{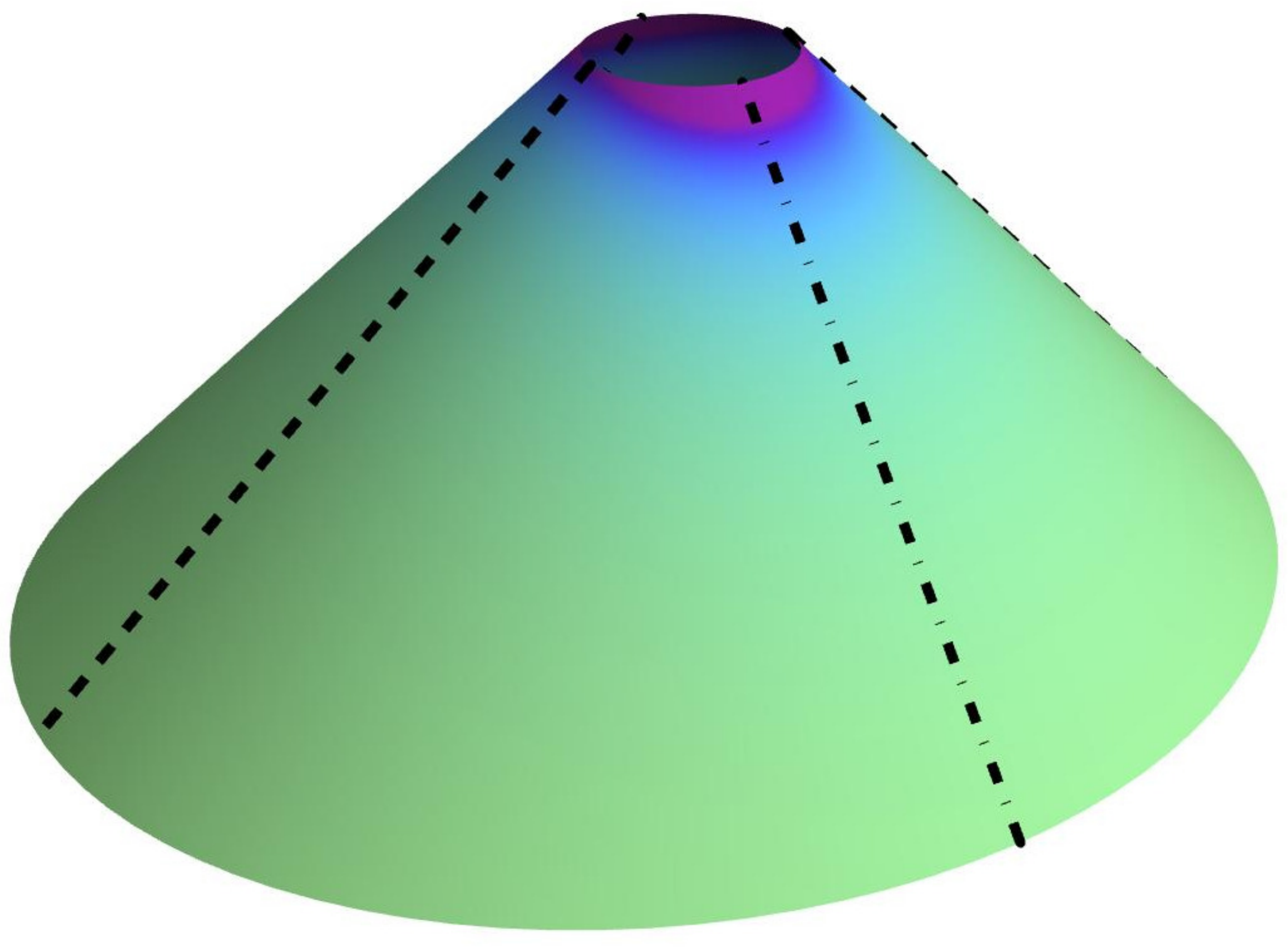}}
\hfil
\subfigure[]{\includegraphics[width=0.32 \textwidth]{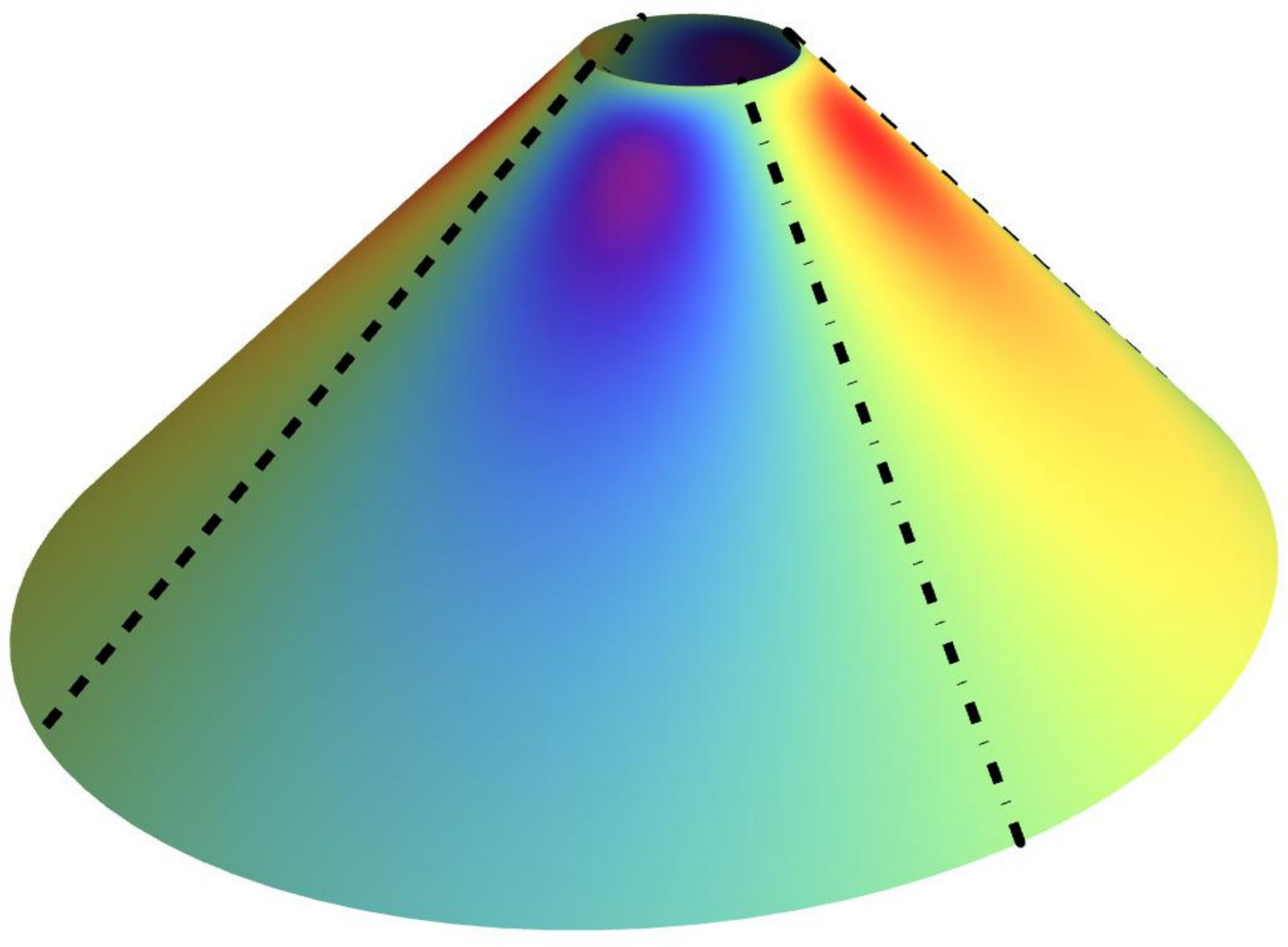}} 
\hfil
\subfigure[]{\includegraphics[width=0.32 \textwidth]{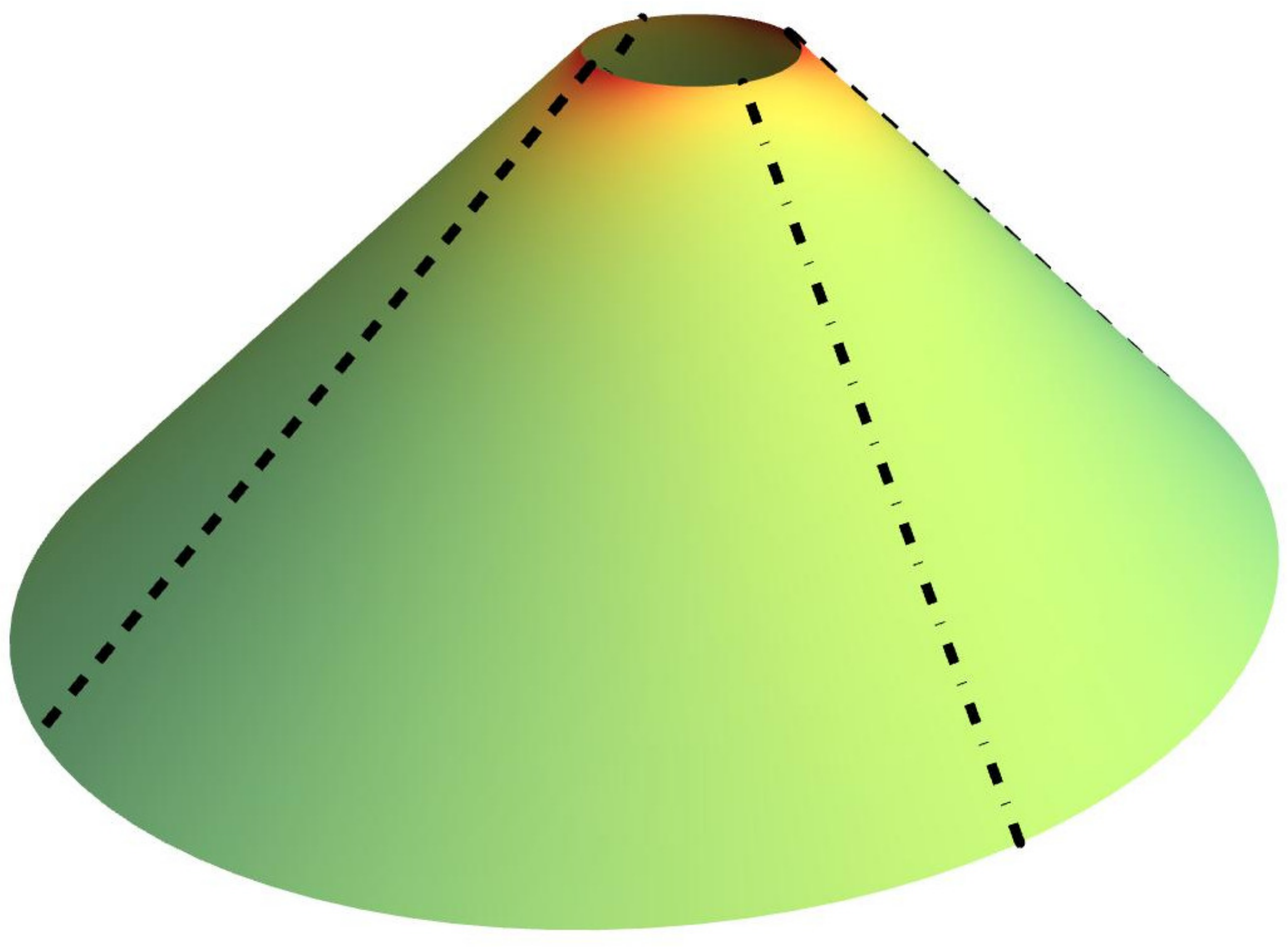}}
\\
\subfigure[$f_{\parallel \parallel}$]{\includegraphics[width=0.32 \textwidth]{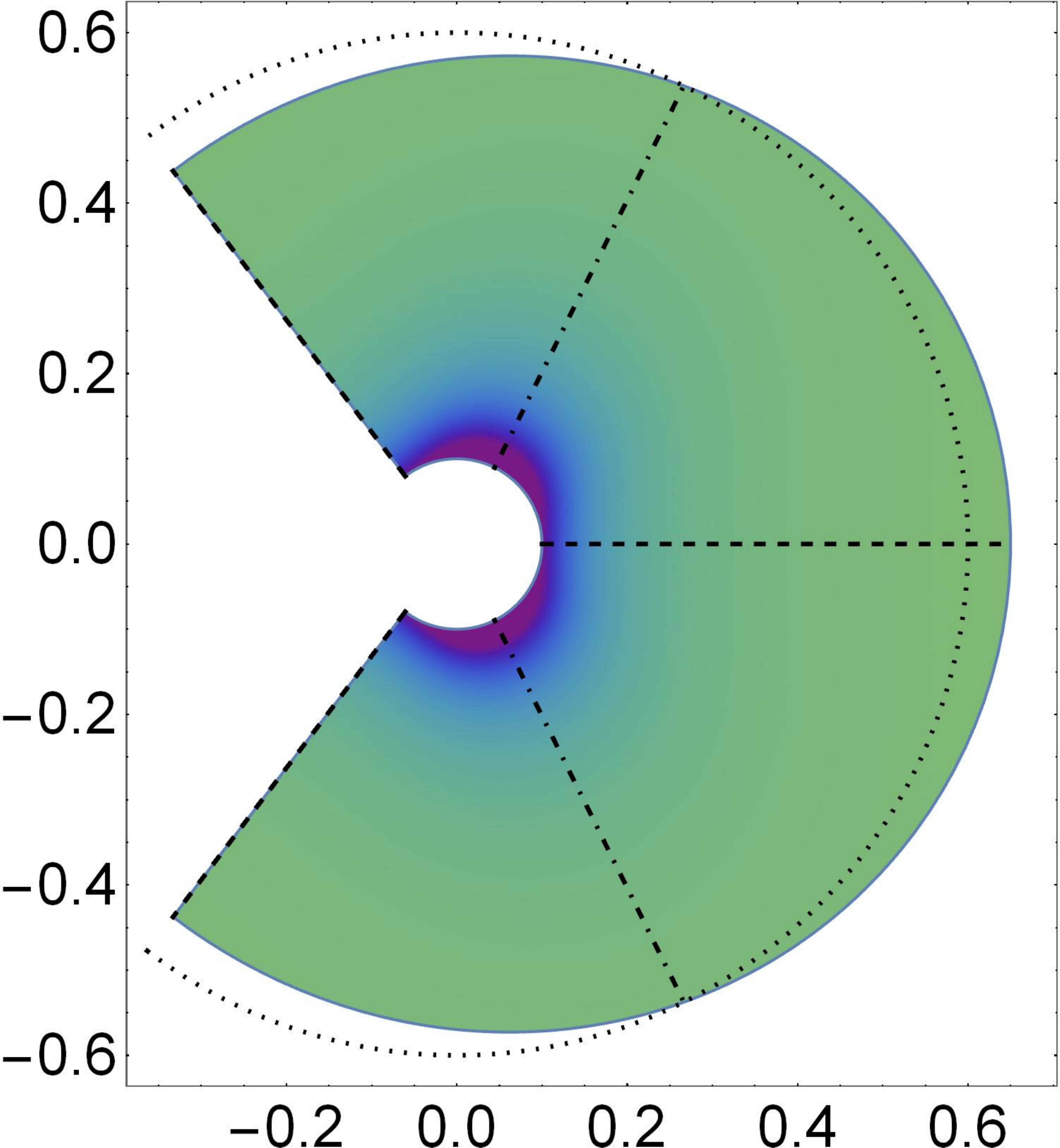}}
\hfil
\subfigure[$f_{\parallel \perp}$]{\includegraphics[width=0.32 \textwidth]{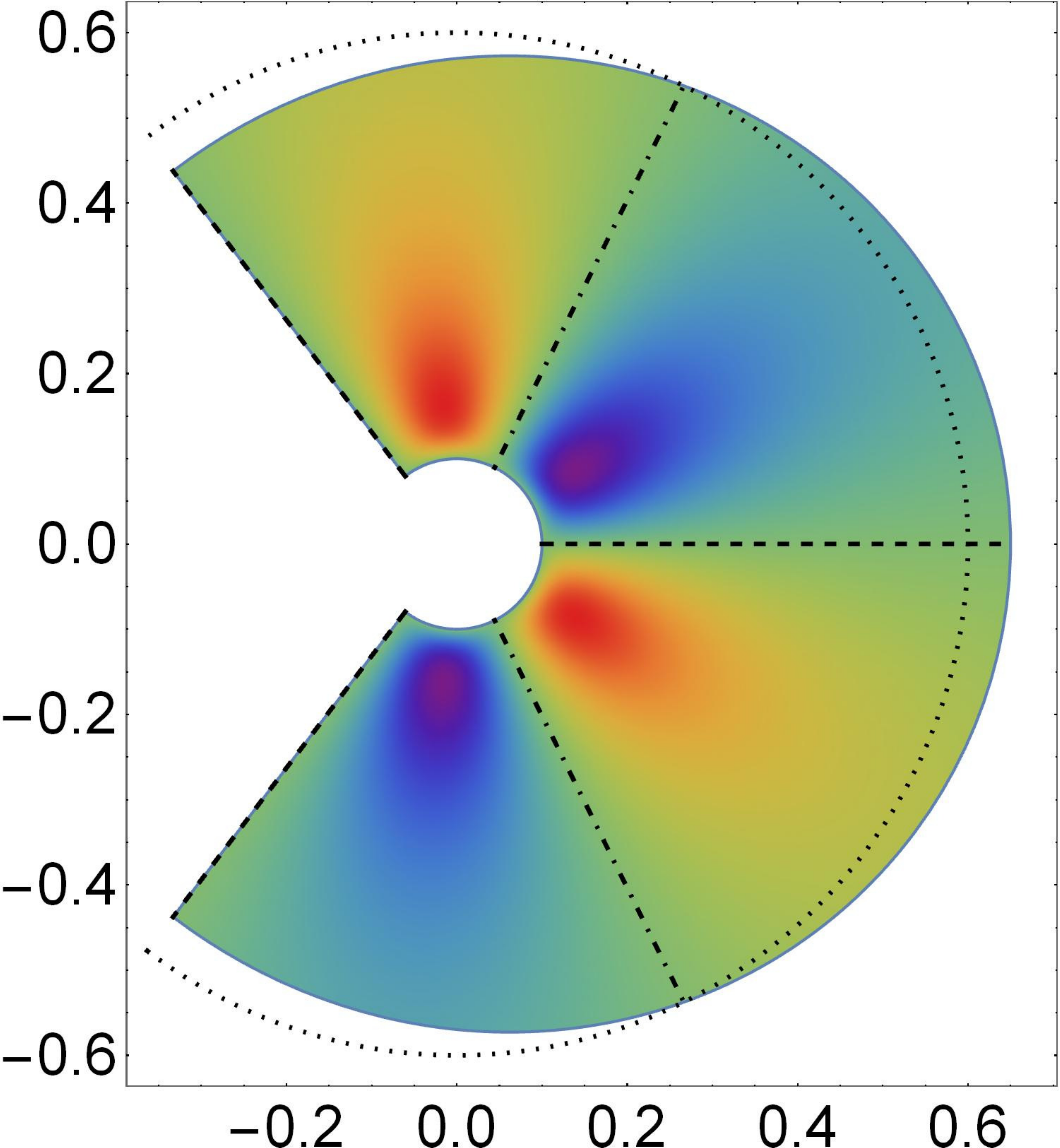}}
\hfil
\subfigure[$f_{\perp \perp}$]{\includegraphics[width=0.32 \textwidth]{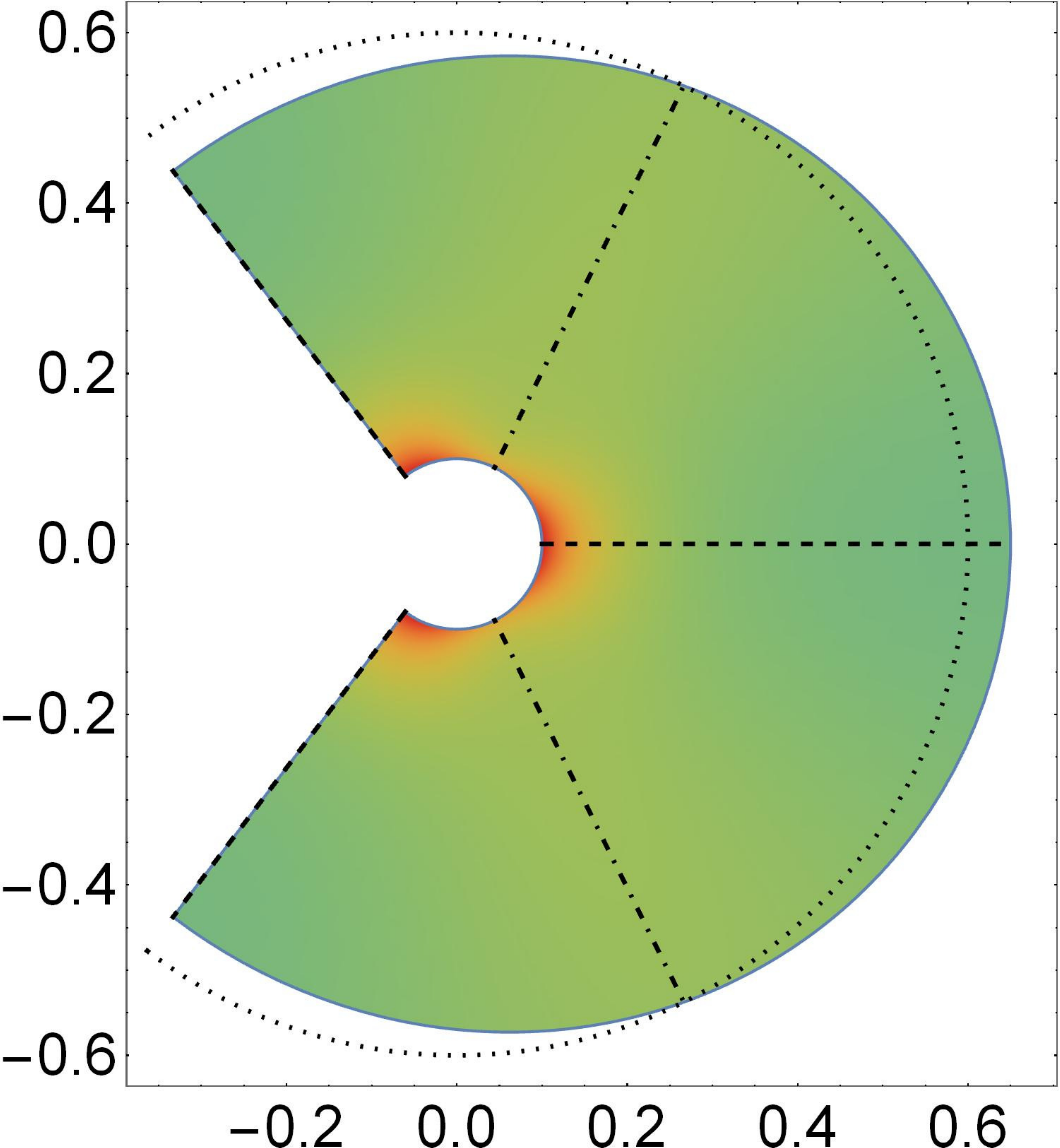}}
\\
\includegraphics[scale=0.6]{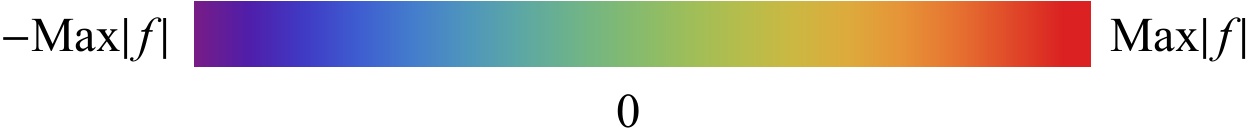}
\caption{(Color online). (a) - (c) Tangential stresses on cones with an elliptic boundary, dashed (dot-dashed) lines represent rays pointing towards the long (short) directions of the elliptic boundary; (d)-(f) tangential stresses shown on a cone that has been \textit{cut} along $\phi_0=\pi$ and laid flat for a better visualization of their distribution, the dotted line represents the original circular boundary. The magnitudes of the stresses have been scaled and color coded: positive (negative) regions indicate compression (tension) and are represented within the range of green-red (green-violet). The magnitudes of the deformation and of the quadrupolar corrections have been exaggerated for illustration purposes. The original cone has $\theta_0=3 \pi/4$ ($\Delta \phi = 2 \pi (1-1/\sqrt{2})$), $r_0=0.1$ and $R_0=0.6$.} \label{fig:conetgstresses}
\end{center}
\end{figure}

In this section we apply these results to a geometry with a boundary that is a small elliptical deformation of a circle. 
\\
The resulting equilibrium geometry will be a deformation of a circular cone, whose generating curve ${\bf u}_0$ is the circle of latitude with $\theta = \theta_0$. 
Let $R(s) = R_0 + R_1 (s)$, so $B = B_0 + B_1(s) + B_2 (s) +\dots$ with $B_0 = \ln (R_0/r_0)>0$, $B_1 = R_1/R_0$ and $B_2 =- B_1^2/2 $. We  expand $\kappa = \kappa_0 + \kappa_1 (s) + \kappa_2 (s) +\dots\;$ and $c$, the constant defined in Eq.(\ref{eq:akappapp}),  $c= c_0 + c_1+ c_2 \dots$. 
\\
At zeroth order, from the first integral (\ref{eq:akappapp}), Eqs. (\ref{phip}) and (\ref{DEtheta}) we get
\begin{equation}
\kappa_0 = - \cot \theta_0\,, \quad \phi_0 = k_0 s\,, \quad c_0 = \frac{B_0}{2} \left(k_0^2+1\right) \,, \quad \mbox{with} \quad k_0^2 = \kappa_0^2 + 1 = \csc^2 \theta_0 \,.
\end{equation}
Recall that $\kappa_0$ is the geodesic curvature of the spherical curve and its normal curvature is unit,\footnote{To get dimensions right recall that ${\bf u}$ is a unit vector.} so $k_0= \csc \theta_0$ is the Frenet Serret curvature.\footnote{The plus (minus) sign of $k_0$ corresponds to a cone opening downwards with $\theta_0 >\pi/2$ (upwards with $\theta_0 < \pi/2$) such that $\bfu$ is traversed anticlockwise (clockwise) about the $Z$ axis.} The arc-length of the generating curve lies within the range $0< s < S_0 = 2\pi/k_0$. Since $B_0$ is constant, the integrability condition (\ref{eq:intcondBkappa}) is satisfied identically at this order.
\\
At first order, the first integral (\ref{eq:akappapp}) reads
\begin{equation} \label{eq:ELeq4}
 B_0 \left( \kappa_1'' + \kappa^2_0 \kappa_1  \right)+ \kappa_0 \left(B_1'' + k_0^2 \, B_1 - \tilde c_1 \right) = 0 \,,
\end{equation}
where $\tilde c_1= \frac{1}{2} \kappa_0^2 b_1 + c_1$.
We consider non-degenerate cones with $\kappa_0\neq 0$ ($\theta_0 \neq \pi/2$). We are interested specifically in an elliptical deformation of the boundary, so we set $R_1 = r_1 \cos \phi_0=r_1 \cos k_0 s$, and thus $B_1=b_1 \cos k_0 s$, with $b_1 = r_1/ R_0$. Notice that it is the Frenet Serret curvature $k_0$, rather than the geodesic curvature, $\kappa_0$, appearing as the argument of the cosine. Now Eq. (\ref{eq:ELeq4}) becomes 
\begin{equation} \label{eq:ELeq5}
B_0 \,( \kappa_1'' + \kappa^2_0 \kappa_1)    - \kappa_0 \tilde c_1  = 0 \,.
\end{equation}
The harmonic solution of Eq. (\ref{eq:ELeq5}) $\sim \cos \kappa_0 s$  is incompatible with periodicity 
($\kappa_0 \neq k_0$). This is reflected in the  integrability condition (\ref{eq:intcondBkappa}).  Whereas, at first order this condition is satisfied identically; at second order it reads $\oint ds (\kappa_0^2 B_2'/2+\kappa_0 \kappa_1 B_1')=0$, implying that $\kappa_1$ should be orthogonal to $B_1'$.  Thus, the only possibility is a constant  
\begin{equation} \label{eq:kappa1c}
\kappa_1 = \frac{\tilde c_1}{B_0 \kappa_0}  \,.
\end{equation} 
To construct the cone, we have for the linearization of Eq.(\ref{DEtheta})
\begin{equation}
\theta _1'' + k_0^2 \, \theta_1 - \kappa_1 = 0\,.
\end{equation}
Integrating this equation, with constant $\kappa_1$, we get
\begin{equation} \label{theta1sol}
\theta_1 = \theta_{1M} \cos k_0 s  + \frac{\kappa_1}{k^2_0}\,,
\end{equation}
where $\theta_{1M}$ is a constant of integration (we have neglected another arbitrary phase).
\\
At first order Eq.(\ref{phip}) reads
\begin{equation}
\phi _1' - \kappa_0 k_0 \theta_1 = 0 \,,
\end{equation}
whose solution, with $\theta_1$ given in Eq. (\ref{theta1sol}), is
\begin{equation} \label{phi1sol}
 \phi_1 =  \kappa_0 \, \left(\theta_{1M} \,\sin k_0 s + \frac{\kappa_1}{k_0} s \right)\,.
\end{equation}
Since $\phi_1(S_0)=  2\pi \kappa_0 \kappa_1 /k_0^2$, any non-vanishing first order correction to the curvature violates the periodicity of the deformation, so $\kappa_1=0$. 
This implies $\tilde c_1=0$ or
$c_1= - \frac{1}{2} \,\kappa_0^2\,b_1/2 $. Therefore, the first order deformations $\phi_1$ and $\theta_1$ are independent of the boundary deformation $R_1$ or $B_1$. 
Consequently there are no first order perturbations of the conical geometry. Modulo 
an infinitesimal rotation of $\bfu_0$ about the $Y$ axis,
\begin{equation}
\mathbf{u}_1 = \theta_1 \hat{\bm{\theta}}_0 + \sin \theta_0 \phi_1 \hat{\bm{\phi}}_0 = \theta_{1M} (\cos \theta_0,0,- \sin \theta_0 \cos \phi_0)\,,
\end{equation}
which we can safely set to zero ($\theta_{1M} = 0$), 
$\theta_1=0$ and $\phi_1=0$. 
\\
The conical geometry first responds to an elliptic deformation of the boundary at second order.
With $\kappa_1=0$,  correct to second order Eq. (\ref{eq:akappapp}) resembles its first order counterpart 
Eq.(\ref{eq:ELeq5}): ($\tilde c_2= \frac{1}{2} \kappa_0^2 b_2 + c_2$)
\begin{equation} \label{eq:1st2ndorder}
B_0 \left(\kappa_2'' + \kappa_0^2 \kappa_2 \right) + \kappa_0 \left( B_2'' + k_0^2 B_2 - \tilde c_2 \right) = 0 \,,
\end{equation}
with the difference that the dipolar $B_1$ gets replaced by a quadrupole, $B_2= b_2 \cos^2 k_0 s$ (with $b_2 = -b_1^2/2$), so that it reads
\begin{equation} \label{eq:1st2ndordersimp}
B_0 \left(\kappa _2'' + \kappa _0^2 \kappa _2\right) + \kappa _0 \left(\frac{k_0^2}{2}  \left(1-3 \cos 2 k_0 s \right) b_2 -\tilde c_2\right) = 0 \,.
\end{equation}
The solution is
\begin{equation} \label{eq:solkappa2}
\kappa_2 = \kappa_{2M} \, \cos 2 k_0 s  + \beta_2 \,,
\end{equation}
where
\begin{equation}
\kappa_{2M} = -\frac{3 \kappa_0 k_0^2 b_2}{2 B_0 \left(3 k_0^2+1\right)} \,, \quad \beta_2 = \frac{1}{B_0 \kappa _0} \left(- \frac{k_0^2}{2} b_2 + \tilde c_2 \right) \,.
\end{equation}
A term $\approx \cos \kappa _0 s$ is again inconsistent with the periodicity of the deformation.   We will see that the constant $\beta_2$ vanishes. The conical curvature responds symmetrically at second order to the deformation: it increases along the direction of the semi-major axis of the conic section, 
$\kappa_2 =\kappa_{2M} >0$ when $s=0$ and $s= S_0/2$; it decreases along the orthogonal direction. One would expect this accidental symmetry along the long direction to be broken at third order.
In any case, we see that the obliquely cut circular cone is no longer circular. 
\\\\
With $\theta_1=0$ and $\kappa_1=0$, at second order Eq. (\ref{DEtheta}) reads  
\begin{equation} \label{eq:2ndordEDTheta} 
\theta _2'' + k_0^2 \theta _2 - \kappa_2 = 0 \,,
\end{equation}
whose solutions for $\kappa_2$ given by Eq.(\ref{eq:solkappa2}) is
\begin{equation} \label{eq:2ndordsolsTheta}
\theta_2 = \theta_{2M}\cos 2 k_0 s + \gamma_2 \cos k_0 s + \frac{\beta_2}{k_0^2}\,, \quad \theta_{2M} =  -\frac{\kappa_{2 M}}{3 k_0^2}\,,
\end{equation}
where $\gamma_2$ is a constant of integration.
\\\\
With $\theta_1=0$ and $\kappa_1=0$, to second order Eq.(\ref{phip}) is given by
\begin{equation} \label{eq:2ndordEDPhi}
\phi _2' - \kappa_0 k_0 \theta_2 =0 \,,
\end{equation}
with solution for $\theta_2$ given in Eq. (\ref{eq:2ndordsolsTheta}) is
\begin{equation} \label{sol:phi2}
\phi_2 = \kappa _0 \left( \frac{\theta_{2 M}}{2} \sin 2 k_0 s + \gamma_2 \sin k_0 s + \frac{\beta_2 }{k_0} s \right)\,.
\end{equation}
As at first order, $\phi_2(S_0)=2 \pi \kappa_0 \beta_2 /k_0^2$ is inconsistent with periodicity unless  $\beta_2=0$ or $\tilde c_2=k_0^2 b_2/2$. 
\\\\
For $\theta_1=0$ and $\phi_1=0$, the second order deformation of the spherical curve is
\begin{eqnarray}
\mathbf{u}_2 &=& \theta_2 \hat{\bm{\theta}}_0 + \sin \theta_0 \phi_2 \hat{\bm{\phi}}_0 \nonumber \\
&=& \theta_{2 M} \left( \cos \theta_0 \cos^3 \phi_0, - \cos \theta _0 \sin ^3 \phi _0 , - \sin \theta _0 \cos 2 \phi _0 \right) \,.
\end{eqnarray}
modulo an irrelevant tilt we ignore. 
\\\\
We now possess the second order deformation of the curvature as well as  the spherical curve describing the cone.  We are now in position to determine the corrections of the stresses on the perturbed cone to this order. From Eq.(\ref{eq:Cparper}) we find that at lowest order we have that $C_{\parallel 0} = c_0 /B_0=(k_0^2+1)/2$, whereas with $c_1= 0$  and $c_2=-k_0^2 b_1^2/4$, we find
\begin{subequations}
\begin{eqnarray}  \label{eq:Cparper1}
C_{\parallel\,1} (s) &=& -\frac{1}{B_0} \left( B_1'' + k_0^2 B_1  \right) = 0\,,\\ 
C_{\parallel\,2} (s)&=& -\frac{1}{B_0} \left(B_2''+k_0^2 B_2 - c_2 + \frac{B_1}{B_0} C_{\parallel 1}(s)\right) = - \frac{3}{4} \frac{k_0^2}{B_0} b_1^2 \cos 2 \phi_0\,.
\end{eqnarray}
\end{subequations}
There is only a second order correction to $C_\parallel$ (and thus to $T_{\parallel \parallel}$).  The correction to the  bending stress is second order because $\kappa_1=0$. 
In \ref{App:pertstress} we evaluate the stress correct to second order by applying perturbation theory to 
the results of section \ref{bstresses}.
We see that there is no first order correction to the remaining components of the tangential stress. 
\\\\
Using the results derived in the appendix, the complete tangential stress, summing the isometry constraining and bending contributions  has components adapted to the unperturbed tangent frame $\{\textbf{t},\textbf{l}\}$ given by 
\begin{subequations}
\begin{eqnarray} \label{eq:fprpcompspert}
f_{\parallel \parallel} &=& \frac{1}{2} \left( \frac{\kappa}{r}\right)^2 + T_{\parallel \parallel}  \approx -\frac{1}{r^2} \left(1-\frac{3 k_0^2}{B_0 \left(3 k_0^2+1\right)} b_1^2  \cos 2 \phi _0 \right)\,, \\
f_{\parallel \perp}  & = & T_{\parallel \perp} \approx -\frac{3 k_0^3 }{2 B_0 r^2} \ln \frac{r}{r_0} \,  b_1^2 \sin 2 \phi_0 \,,\\
f_{\perp \perp} & = & -\frac{1}{2} \left( \frac{\kappa}{r}\right)^2 + T_{\perp \perp} \approx \frac{1}{r^2} \left[1 + \frac{3 k_0^4}{B_0} \left(\ln \frac{r}{r_0} - \frac{r}{R_0} \left(\frac{3 k_0^2}{3 k_0^2+1} + 2B_0 \right)+\frac{3 k_0^2}{3 k_0^2+1}\right) b_1^2 \cos 2 \phi_0 \right] \, .
\end{eqnarray}
\end{subequations}
As remarked in Sec. \ref{stresscone}, to lowest order these tangential stresses on a cone with circular boundary do not depend on the deficit angle $\Delta \varphi$, but we do see that their second order corrections on a cone with an elliptic boundary depend on $\Delta \varphi$ through the Frenet Serret curvature $k_0 = \csc \theta_0 = 2\pi/(2 \pi -|\Delta \varphi|)$. These tangential stresses of cones with elliptic boundaries are plotted in Fig. (\ref{fig:conetgstresses}). A positive (negative) sign of its components indicates that the surface is under compression (tension). These tangential stresses are concentrated near the apex and decay as $r^{-2}$. 
\vskip1pc \noindent
$f_{\parallel \parallel}$ is  the tangential force per unit length exerted along the positive azimuthal direction across rays of constant $s_c$ by the region with $s < s_c$ on that with  $s>s_c$. Since $f_{\parallel \parallel}<0$, the cone is under lateral tension. The quadrupolar correction correlates with the correction to the curvature $f_{\parallel \parallel 2} \propto \kappa_2$. It  decreases (increases) the magnitude of the lateral tension along the two rays pointing towards the long (short) direction of the elliptic boundary, see Figs. (\ref{fig:conetgstresses}a) and (\ref{fig:conetgstresses}d). At this order the local stresses do not provide a distinction between the two rays pointing towards the long direction of the elliptic boundary, even though their lengths differ ($R_0 \pm r_1$). However, evaluating the total force along the two, $F_{\parallel \parallel} = \int dr f_{\parallel \parallel}$ at this order, we find that the shorter ray is under greater total tension, for $F_{\parallel \parallel short}- F_{\parallel \parallel long} \propto 2 r_1/(R^2_0-r^2_1)$, consistent with expectations. The tension along the two short directions with equal unchanged ray length is also increased, a non-linear correction that is less intuitively obvious. 
\vskip1pc \noindent
The off-diagonal component $f_{\parallel \perp}$ is the magnitude of the shear force per unit length exerted along the radial direction across a ray of constant arc length $s_c$, by a region with $s < s_c$
on the region with $s > s_c$;  this is equal to the shear force along the azimuthal direction exerted by a region bounded by a line of constant $r_c$ on the region with $r<r_c$.  At lowest order, $f_{\parallel \perp}$ vanishes, so it is given entirely by the quadrupolar correction, which correlates with the derivative of the correction to the curvature $f_{\parallel \perp 2} \propto \kappa'_2$. Thus, this shearing force vanishes along the  four rays pointing towards the long and short directions of the elliptic boundary. One would expect this by symmetry. Its magnitude is a maximum along the four diagonal directions. The shear force exerted across the diagonals rays, long or short, towards the long  elliptic directions is 
outwards with the same value; whereas that towards the short directions is inwards as confirmed by Figs. (\ref{fig:conetgstresses}b) and (\ref{fig:conetgstresses}e).  One would expect the symmetry under  reflection $\phi_0 \rightarrow -\phi_0$ to persist at all orders;  the quadrupole approximation implies 
symmetry under interchange $\phi_0 \rightarrow 2\pi - \phi_0$ as well.
At second order, we fail to distinguish between the rays pointing towards the long and the short diagonals. This accidental symmetry is broken only at third order in perturbation theory.
\\\\
$f_{\perp \perp}$ is the magnitude of the tangential force per unit length along the radial direction exerted by a region bounded by a line of constant $r_c$ on the region with $r<r_c$. Since $f_{\perp \perp}>0$, the cone is under radial compression. Like $f_{\parallel \parallel}$, the second order correction to $f_{\perp \perp}$ correlates with the correction to the curvature $f_{\perp \perp 2} \propto \kappa_2$. But this time the correction increases (decreases) the magnitude of the radial compression along the rays pointing towards the long (short) direction of the elliptic boundary, see Figs. (\ref{fig:conetgstresses}c) and (\ref{fig:conetgstresses}f). Calculating the total force along a closed circle of constant $r$, we see that the quadrupolar correction does not contribute: $F_{\perp \perp} = \int ds f_{\perp \perp} \approx 2\pi/(k_0 r_c^2)$.


\section{Conclusions}

We have presented a framework allowing us to analyze the stress distribution in  thin-sheets that undergo isometric bending without requiring  the introduction of boundary layers. This involves the introduction of two local constraints on the boundary curve, one fixing its arc-length; and less obviously, another fixing its geodesic curvature with respect to the sheet which it bounds. Together they ensure that boundary deformations are consistent with isometry. We have applied this framework to a cone, isometric to a circular wedge, with a deficit angle and also, but to a lesser extent, with a surplus angle. The equilibrium shapes in the absence of external forces are remarkably insensitive to boundary conditions; indeed they can be determined without even knowing what the correct conditions are. The stress distributed within the cone, on the other hand, does depend sensitively on these conditions.   
In a circular paper cone we showed that the stress is diagonal with respect to the radial and circular directions (in compression along the former, tension along the latter loops), falling off quadratically with distance from the apex. The existence of this tension is not surprising; the independence of its magnitude on the deficit angle is.
\vskip1pc \noindent
If the cone is  cut from a non-circular wedge, and in particular one forming an elliptical boundary, 
the determination of the geometry no longer decouples from the boundary. We have solved the EL equations in the bulk perturbatively.  
The bulk geometry surprisingly only depends on the elliptical boundary conditions at second order in its eccentricity.  
The conical curvature increases equally at it short and the long ends; it is reduced in the orthogonal direction along which the distance to the apex is unchanged. There is also no first order change in the stress distributed throughout the cone. The quadrupolar change in the stress implies that the diagonal stresses (tension along loops, compression along the radial direction) are changed: the lateral tension is no longer equal in magnitude to the radial compression; both changes depend sensitively on the deficit angle.  The angular dependence of the two as well as the radial falloff differ. The radial stress, is in fact dominated by its correction remote from the apex with an inverse $r$ falloff; the exact quadratic falloff of the lateral tension is, however, preserved. With the perturbed boundary, the principal stresses no longer coincide with the principal directions: with respect to the latter directions shearing stresses come into play. 
\vskip1pc \noindent
The decision to focus on paper cones, despite appearances, was not a frivolous one; obtaining reliable results has not come easily. This framework can clearly be extended to other geometries where isometry is a relevant constraint.  Suppose, instead of a conical sheet with an apex, we form an annulus by excising a finite disc at its center.  When there is a surplus angle, we showed in a previous paper that the only stable equilibrium of the corresponding cone is the ground state with a two-fold symmetry \cite{GMV12}. One might \textit{hope} that this is also the ground state of the corresponding annulus. But this would require the rulings on the annulus to intersect  at  the phantom apex that organized the cone. In fact, it is simple to show that such a  state is not even stationary never mind it forming the ground state; for it is impossible to satisfy the boundary conditions on the new inner boundary if the outer boundary is finite.  
Indeed,  it can be shown that there  are tangent-developable deformations of a surplus conical annulus lowering its energy. The upshot: the conical annulus with two-fold dihedral symmetry is not the ground state. Its rulings do not converge on a single point.  How it does this appears also to involve symmetry breaking. The construction of this surface is a challenge for the future.

\section*{Acknowledgements}

We have benefited from discussions  with Eliot Fried, James Hanna,  Gert Van Der Heijden, and David Steigmann.
JG was supported in part by CONACyT grant No. 180901, PVM was supported by Cátedras CONACyT grant No. 439-2018.

\begin{appendix}

\setcounter{equation}{0}
\renewcommand{\thesection}{Appendix \Alph{section}}
\renewcommand{\thesubsection}{A. \arabic{subsection}}
\renewcommand{\theequation}{A.\arabic{equation}}

\section{Derivation of the change of the boundary energy} \label{App:dHbdry}

The change in the energy due to a variation of the boundary, given in Eq.(\ref{delHCB}) can be recast as
\begin{equation} \label{dHbdryDBX}
\delta H_{C \,{\sf boundary}} = - \int d \ell \left( \bff_\perp \cdot \delta \bfX - (H_{\parallel \perp} \bfT + H_{\perp \perp} \bfL) \cdot \delta \bfn \right)\,,
\end{equation}
where 
\begin{subequations} \label{fprpHparprpHprpprp}
\begin{eqnarray} 
\bff_\perp &=& \left( T_{\parallel \perp} -H \tau_g \right) \bfT + \left( T_{\perp \perp} + \frac{K}{2} \left( \kappa_{n \perp} -\kappa_n \right) \right) \bfL -\nabla_\perp K \bfn\,,\\
H_{\parallel \perp} &=& \rmL_a \rmT_b H^{ab} = \mathrm{k}_G \tau_g \,, \\
H_{\perp \perp} &=& \rmL_a \rmL_b H^{ab} = K + \mathrm{k}_G \kappa_n \,,
\end{eqnarray}
\end{subequations}
and $H^{ab}$ is given by Eq.(\ref{Habdef}).
From the variation of the embedding functions (\ref{delXtn}), we have that the variation of the tangent and normal vectors are 
\begin{subequations}
\begin{eqnarray} \label{eq:deltaeadeltan}
\delta \bfe_a &=& \nabla_a \delta \bfX = \left( \nabla_a \Psi^b + K_{a}^{\phantom{a}b} \Phi \right) \bfe_b + \left( \nabla_a \Phi - K_{ab} \Psi^b \right) \bfn \,,\\
\delta \bfn &=& - \bfn \cdot \delta \bfe_a g^{ab} \bfe_b = - \left( \nabla_a \Phi - K_{ab} \Psi^b \right) \bfe_b\,.
\end{eqnarray}
\end{subequations}
The variation of the normal vector can be spanned in the boundary Darboux frame as
\begin{equation} \label{deltanDrbxbdr}
\delta \bfn = - \left( \dot{\Phi} - \kappa_n \Psi_\parallel + \tau_g \Psi_\perp\right) \bfT - \left(\nabla_\perp \Phi + \tau_g \Psi_\parallel - \kappa_{n \perp} \Psi_\perp \right) \bfL \,.
\end{equation}
Substituting Eqs. (\ref{fprpHparprpHprpprp}) and (\ref{deltanDrbxbdr}) in Eq.(\ref{dHbdryDBX}), and integrating by parts the term involving $\dot{\Psi}$, we obtain Eq.(\ref{boundaryvar}).

\setcounter{equation}{0}
\renewcommand{\thesection}{Appendix \Alph{section}}
\renewcommand{\thesubsection}{B. \arabic{subsection}}
\renewcommand{\theequation}{B.\arabic{equation}}

\section{Derivation of the variation of the boundary geodesic curvature} \label{App:derkappag}

Here we present a derivation of Eq.(\ref{delkg}) that does not make use of redundant or inappropriate information along the boundary curve. Specifically, we avoid making use of the full metric variation
for the reasons explained in the text.
\vskip1pc \noindent
We first recall that, for any scalar function $F$ \cite{CCG}
\begin{equation} \label{delperps}
\delta_\perp (F') = -\frac{ \delta_\perp ds}{ds} \, F' + 
(\delta_\perp F)' =
(\kappa_g \Psi_\perp  - \kappa_n \Phi) \, F' + (\delta_\perp F)' \,.
\end{equation}
In particular,  if $F$ coincides with one of the Cartesian embedding functions $F=Y^i$, then Eq.(\ref{delperps}) implies that
\begin{eqnarray}
\delta_\perp \bfT &=& ( \kappa_g \Psi_\perp  - \kappa_n \Phi) \, \bfT +
(\Psi_\perp \bfL + \Phi\,\bfn)'\nonumber\\
 &=& (\Psi'_\perp  - \tau_g \Phi) \bfL + (\Phi' + \tau_g \Psi_\perp) \bfn \,.
\end{eqnarray}
Notice that the term proportional to $\bfT$ vanishes, not due to isometry (Eq.(\ref{delgproj}a)), but to the fact that $\bfT$ is unit.
Repeating this argument for the scalars $F=t^i$, we obtain upon projection along $\bfL$:
\begin{equation}
\bfL \cdot  \delta_\perp (\bfT') = \kappa_g ( \kappa_g \Psi_\perp - \kappa_n \Phi) + 
(\Psi'_\perp  - \tau_g \Phi)' - \tau_g  (\Phi' + \tau_g \Psi_\perp )
\end{equation}
The corresponding change in the geodesic curvature is given by
\begin{equation}
\delta_\perp \kappa_g = \bfL \cdot \delta_\perp (\bfT') + \delta_\perp \bfL \cdot \bfT' 
=\bfL \cdot \delta_\perp (\bfT') + \kappa_n \bfL \cdot \delta_\perp \bfn
\end{equation}
To express $\bfL\cdot \delta_\perp \bfn$ in terms of the projections, we note that  
\begin{eqnarray} 
\bfL \cdot \delta_\perp \bfn &=& -\bfn \cdot \nabla_\perp (\Psi_\perp \bfL + \Phi\,
\bfn) \nonumber\\
&=& - \mathbf{n} \cdot \nabla_\perp \bfL \, \Psi_\perp - \nabla_\perp\Phi\,.
\end{eqnarray}
But $\bfn \cdot \nabla_\perp \bfL = - K_{ab} \, \rmL^a \rmL^b = - \kappa_{n\perp}$.
Thus
\begin{equation}
\delta_\perp \kappa_g = \Psi''_\perp + (\kappa_g^2 + \kappa_n \kappa_{n\perp} - 
\tau_g^2 ) \Psi_\perp   - \kappa_g  \kappa_n \Phi
- (\tau_g \Phi)' - \tau_g \Phi'  - \kappa_n \nabla_\perp \Phi \,.
\end{equation}
Finally using the expression of the Gaussian curvature in terms of the Darboux curvatures
\begin{equation} \label{KGkappa}
K_G = \kappa_n \kappa_{n \perp} -\tau_g ^2\,,
\end{equation}
we reproduce Eq.(\ref{delkg}).
\vskip1pc \noindent
An alternative  derivation of  Eq.(\ref{delkg}) proceeds from the intrinsic definition of $\kappa_g$.
However, this derivation uses the metric variation explicitly and with it, terms proportional to $\nabla_\perp \Psi_\parallel$ and $\nabla_\perp \Psi_\perp$, present in Eqs. (\ref{delgproj2}) and (\ref{delgproj3}) even if they do conspire to cancel in the calculation. It may be of interest to perform the variation this way in order to pinpoint where these cancellations occur in the calculation.  


\setcounter{equation}{0}
\renewcommand{\thesection}{Appendix \Alph{section}}
\renewcommand{\thesubsection}{C. \arabic{subsection}}
\renewcommand{\theequation}{C.\arabic{equation}}

\section{Force Balance} 
\label{Force Balance}

Suppose that $\delta \bfX = \delta \mathbf{a}$, a constant vector on the boundary, such that $\Psi_\parallel = \delta \bf{a} \cdot \bfT$, $\Psi_\perp = \delta \bf{a} \cdot \bfL$ and $\Phi = \delta \bf{a} \cdot \bfn$. From Eq.(\ref{boundaryvar}), we have that in equilibrium, the change in the energy is given by the boundary term  
\begin{equation} \label{f1}
\delta H_{C\,{\sf boundary}} = - \delta \mathbf{a} \cdot \int d \ell \, \left[   (T_{\parallel \perp}  - K \tau_g) \bfT + 
\Big(T_{\perp\perp} + \frac{1}{2} ( \kappa_{n\perp}^2 - \kappa_n^2 ) \Big) \bfL  - \nabla_\perp K\, \mathbf{n} \right]\,.
\end{equation}
Terms proportional to $\mathrm{k}_G$ can be shown to vanish on account of the Darboux Eq. for $\bfn$ (\ref{BndyDbxeqs}), the identity $\nabla_\perp \bfn = L^a \nabla_a \bfn = -\tau_g \bfT + \kappa_{n\perp} \bfL$ and the expression of $K_G$ given in Eq.(\ref{KGkappa}).
\\
Modulo the boundary conditions (\ref{bcs}), the identity (\ref{KGkappa}), the integrand of Eq.(\ref{f1}) and integrating by parts,  it can be cast as a total derivative: 
\begin{equation} \label{f2}
\delta H_{C\,{\sf boundary}} = - \delta \mathbf{a} \cdot \int d\ell \, \left(- {\cal T} \, \bfT  
 + \dot{\Lambda} \, \bfL + \Lambda \, \tau_g \, \bfn \right) \dot{}\, \,.
\end{equation}
This integrand does not vanish locally but it integrates to zero if the boundary is closed. However, on taking integrations by parts to identify the independent variations in the boundary conditions, we subtracted boundary terms which break the Euclidean invariance of $\delta (H_1 + H_2)$. The omitted terms are
\begin{subequations} \label{eqs:bdrcnst}
\begin{eqnarray}
\delta H_1  &=&  \int d \ell \, ( {\cal T} \Psi_\parallel) \,\dot{} \\
\delta H_2 &=& \int d \ell \,  ( \Lambda \kappa_g \Psi_\parallel + \Lambda \dot{\Psi}_\perp - \dot{\Lambda} \Psi_\perp  - 2 \Lambda\tau_g \Phi) \, \dot{}\,\,,
\end{eqnarray}
\end{subequations}
Not surprisingly, for a translation they coincide with the expression for $\delta H_{C\,{\sf boundary}}$ given by Eq.(\ref{f2}).  
Thus  boundary constraints are also important to balance forces across the free boundary.
\vskip1pc \noindent
For a boundary curve of constant $r=R$ on a cone, $\kappa_g = 1/R$, $\kappa_n= \kappa/R$ and $\tau_g=0$, $\Lambda =-1$ and ${\cal T}=-1/R$ 
\begin{subequations} \label{eqs:bdrcnsttrns}
\begin{eqnarray}
\delta H_1 &=& -\frac{1}{R} \delta \mathbf{a} \cdot \int d \ell\, \dot{\bfT}\,, \qquad \dot{\bfT} = \frac{1}{R} \left( \bfL - \kappa \,\bfn \right)\,,\\
\delta H_2 &=& 0\,.
\end{eqnarray}
\end{subequations}
The force transmitted across the boundary is given by
\begin{equation}
\mathbf{F} = \oint d \ell \, \mathbf{f}_\perp \,,
\end{equation}
where $\mathbf{f}_\perp$ has a bulk contribution,
given by
\begin{equation}
\mathbf{f}_\perp =  T_{\parallel \perp} \bfT + \left(T_{\perp\perp} -\frac{\kappa^2}{2 R^2} \right)\bfL + \frac{\kappa}{R^2} \bfn  \,.
\end{equation}
On the boundary $T_{\parallel \perp}=0$ and $T_{\perp\perp} = 1/R^2 \left(
\kappa^2/2 +1\right)$  ($C_\parallel$ constant), so $\bff_\perp$ is given by
\begin{equation}
\mathbf{f}_\perp = \frac{1}{R} \dot{\bfT}\,.
\end{equation}
This is canceled by an equal contribution from the boundary constraints (\ref{eqs:bdrcnsttrns}).
Thus the local force across the boundary vanishes. Note that the bulk and boundary contribution are each proportional to the derivative $\bfT'$. We have already seen how $H_{2}$ contributes to local torque balance. It would appear that this treatment of forces on the boundary is related to the work in \cite{Aharoni17}.

\setcounter{equation}{0}
\renewcommand{\thesection}{Appendix \Alph{section}}
\renewcommand{\thesubsection}{D. \arabic{subsection}}
\renewcommand{\theequation}{D.\arabic{equation}}

\section{Perturbative Evaluation of the Stress in a Cone with oblique boundary} 
\label{App:pertstress}

Expand $T_{\parallel \perp}^{\calD} = T_{\parallel \perp}^{\calD}{}_0 + T_{\parallel \perp}^{\calD}{}_1 + \dots $, $T_{\perp \perp}^{\calD} = T_{\perp \perp}^{\calD}{}_0 + T_{\perp \perp}^{\calD}{}_1 + \dots$, $C_{\parallel \perp} = C_{\parallel \perp 0} + C_{\parallel \perp 1}+ \dots$ and $C_{\perp} = C_{\perp 0} + C_{\perp 1}+ \dots$. 
At zeroth order the isometric tangential boundary stresses are given by
\begin{subequations}
\begin{eqnarray}
T_{\parallel \perp}^{\calD}{}_0 & = & 0 \,,  \\
T_{\perp \perp}^{\calD}{}_0 & = & \frac{k_0^2+1 }{2 \, R_0^2} \,, \\
C_{\parallel \perp 0} & = & 0 \,, \\
C_{\perp 0} & = & 0\,,
\end{eqnarray}
\end{subequations}
consistent with Eqs. (\ref{TparprpTprpbvar}) and (\ref{CparprpCprpbvar}).
At  first order we obtain
\begin{subequations}
\begin{eqnarray}
T_{\parallel \perp}^{\calD}{}_1 & = & \frac{B_1^{'''}- B_1'}{R_0^2} = k_0 (k_0^2 + 1) \frac{b_1}{R_0^2} \, \sin k_0 s \,, \\
T_{\perp \perp}^{\calD}{}_1 & = & -\frac{k_0^2 + 1}{R_0^2} B_1\,, \\
C_{\parallel \perp 1} & = & \left( \frac{\ln R_0}{B_0} -1 \right) \left(B_1^{'''} + k_0^2 B_1'\right) = 0 \,, \\
C_{\perp 1} & = & \frac{1}{B_0 R_0} \left((B_0+1) B_1^{''''}+\left((B_0 + 1) k_0^2 + 1\right) B_1'' + k_0^2 B_1 \right) = 0 \,.
\end{eqnarray}
\end{subequations}
The second order corrections are
\begin{subequations}
\begin{eqnarray}
T_{\parallel \perp}^{\calD}{}_2 & = &- \frac{3}{2} \frac{k_0}{R_0^2} \left(2 k_0^2+1\right) b_1^2 \sin 2 k_0 s\,, \\
T_{\perp \perp}^{\calD}{}_2 & = &\frac{b_1^2}{4 R_0^2} \left(\left(\frac{3 \left(k_0^2-1\right) k_0^2}{B_0 \left(3 k_0^2+1\right)} - 10 k_0^4+5 k_0^2+3\right) \cos 2 k_0 s - 2 k_0^4 + k_0^2+3\right) \,,\\
C_{\parallel \perp 2} & = & \frac{3 b_1^2 k_0^3 }{2 B_0} \ln r_0 \sin 2 k_0 s\,, \\
C_{\perp 2} & = & -\frac{3 b_1^2 k_0^4}{R_0} \left(\frac{3 k_0^2}{B_0 \left(3 k_0^2+1\right)}+2\right) \cos 2 k_0 s\,.
\end{eqnarray}
\end{subequations}
Although the boundary isometric tangential stresses have first order corrections, their bulk counterparts do not; their second order corrections are 
\begin{subequations}
\begin{eqnarray}
T_{\parallel \perp 0} & = & 0 \,, \\
T_{\perp \perp 0} & = & \frac{k_0^2+1}{2 r^2} \,, \\
T_{\parallel \perp 1} & = & \frac{1}{r^2} \left(B_1^{'''} + k_0^2  B_1' \right) = 0 \,, \\
T_{\perp \perp 1} & = & \left(- \frac{1}{r^2} + \frac{1}{r R_0}\right) \left(B_1^{''''} + \kappa_0^2 B_1''\right)  = 0\,, \\
T_{\parallel \perp 2} & = & -\frac{3 k_0^3 }{2 B_0 r^2} \ln \left( \frac{r}{r_0} \right)  b_1^2 \sin 2 k_0 s\,, \\
T_{\perp \perp 2} & = & \frac{3 k_0^2}{B_0}  \left(\frac{1}{r^2} \left(k_0^2  \left( \ln \frac{r}{r_0} +1\right) - \frac{1}{4} \right)-\frac{k_0^2}{r R_0} \left(2 B_0 + \frac{3 k_0 ^2}{3 k_0^2+1}\right) \right) b_1^2 \cos 2 k_0 s \,.
\end{eqnarray}
\end{subequations}
We thus see that the (first order) elliptic deformation of the boundary induces (second order) two-fold symmetric deformations of $\bf{u}$ and of the tangential stresses. Note that the cutoff length $r_0$  is treated as small compared to $R_0$ but finite in order to maintain the legitimacy of this expansion.

\end{appendix}


\end{document}